\documentclass[11pt]{article}

\usepackage[margin=0.9in]{geometry}
\usepackage{amsmath,amssymb,amsthm,mathtools}
\usepackage{microtype}
\usepackage{enumitem}
\usepackage[bottom]{footmisc}

\usepackage[T1]{fontenc}
\usepackage[utf8]{inputenc}

\usepackage{newtxtext}
\usepackage{newtxmath}
\usepackage{eucal}

\usepackage{bm}

\usepackage{graphicx}
\graphicspath{{figures/}}
\usepackage{booktabs}
\usepackage{array}
\usepackage{tabularx}

\usepackage{xcolor}

\setlist{leftmargin=*,itemsep=0.25em,topsep=0.35em}
\renewcommand{\arraystretch}{1.12}
\allowdisplaybreaks

\usepackage[
  hypertexnames=false,
  colorlinks=true,
  linkcolor=blue,
  citecolor=blue!50!black,
  urlcolor=blue
]{hyperref}
\newcommand{\papertitle}{Classification Aggregation with Multiple Memberships}
\newcommand{\paperauthor}{Federico Fioravanti, Leo Kurata, Vasileios Stamatis}

\AtBeginDocument{\hypersetup{
    pdftitle={\papertitle},
    pdfauthor={\paperauthor},
    pdfsubject={Classification Aggregation with Multiple Memberships}
}}

\usepackage{url}
\usepackage{natbib}
\setcitestyle{authoryear,round,citesep={;}}
\usepackage[hyperpageref]{backref}
\AtBeginDocument{}

\renewcommand*{\backrefalt}[4]{\ifcase #1
  \else
  \space [#2]\fi
}

\newcommand{\N}{\mathbb{N}}

\newcommand{\cl}{\bm c}

\newcommand{\C}{\mathcal{C}}
\newcommand{\Cbar}{\bar{\mathcal{C}}}

\newcommand{\U}{\mathcal U}
\newcommand{\G}{\mathcal G}
\newcommand{\F}{\mathcal F}
\newcommand{\I}{\mathcal I}
\newcommand{\W}{\mathcal W}

\theoremstyle{definition}
\newtheorem{definition}{Definition}

\newtheoremstyle{mystyle}{}{}{\upshape}{}{\bfseries}{.}{ }{\thmname{#1}\thmnumber{ #2}\thmnote{ (#3)}}
\theoremstyle{mystyle}
\newtheorem*{axiom}{Axiom}

\theoremstyle{plain}
\newtheorem{lemma}{Lemma}
\newtheorem{proposition}{Proposition}
\newtheorem{theorem}{Theorem}
\newtheorem{corollary}{Corollary}
\newtheorem{claim}{Claim}
\newtheorem{fact}{Fact}

\theoremstyle{remark}

\renewcommand{\qedsymbol}{$\blacksquare$}

\usepackage{thmtools}
\declaretheorem[
  style=definition,
  name=Example
]{example}

\renewcommand{\thmcontinues}[1]{Continued}

\usepackage{caption}
\usepackage{cleveref}
\crefname{theorem}{Theorem}{Theorems}
\crefname{proposition}{Proposition}{Propositions}
\crefname{lemma}{Lemma}{Lemmata}
\crefname{corollary}{Corollary}{Corollaries}
\crefname{claim}{Claim}{Claims}
\crefname{fact}{{Fact}}{{Facts}}
\crefname{remark}{Remark}{Remarks}
\crefname{example}{{Example}}{{Examples}}
\crefname{appendix}{Appendix}{Appendices}

\crefname{equation}{eq.}{eqs.}
\Crefname{equation}{Equation}{Equations}
\crefname{align}{eq.}{eqs.}
\Crefname{align}{Equation}{Equations}
\crefname{section}{Section}{Sections}

\crefname{table}{Table}{Tables}

\usepackage{chngcntr}
\counterwithin{equation}{section}

\usepackage{autonum}

\usepackage{authblk}

\title{\papertitle}
\author[1]{Federico Fioravanti}
\author[2]{Leo Kurata}
\author[3,4]{Vasilis Stamatis}
\affil[1]{Universit\'{e} Jean Monnet Saint-\'{E}tienne, CNRS, Universit\'{e} Lyon 2, emlyon business school, GATE, 42023, Saint-\'{E}tienne, France.}
\affil[2]{Graduate School of Economics, Waseda University, Tokyo, Japan}
\affil[3]{University of West Attica, Athens, Greece}
\affil[4]{LAMSADE, Universit\'{e} Paris Dauphine--PSL, Paris, France}
\date{November 2026}

\begin{document}

\maketitle

\begin{abstract}
  \noindent
People face the problem of classifying objects into categories while aggregating their opinions. We axiomatically study such a classification aggregation problem with multiple memberships, meaning that each individual assigns one category to every object, whereas the resulting classification may assign more than one category.
Under the natural settings and axioms (i.e., Independence and Supportiveness), each of our new axioms -- Responsibility, Split Stability and Rejection Stability -- characterises what we call the junta rules.
The rules assign a category to an object exactly when some member of a fixed coalition does so.
The coalitions that can secure an assignment form a grill, a topological structure implied by an ultrafilter.
Hence, the study parallels the literature on Arrow's theorem characterised by the decisive coalition structures.
An infinite population setting is also investigated to clarify the differences among the axioms. 

  \medskip
  \noindent\textit{Keywords}: classification aggregation, multiple memberships, junta rules, grills, filters, infinite populations

  \noindent\textit{JEL classification}: C02, C71, D71
\end{abstract}

\begingroup
\makeatletter
\renewcommand{\@makefntext}[1]{\parindent 1em
  \noindent\hb@xt@1.8em{\hss}\hspace{0em}#1}
\footnotetext{
  Fioravanti acknowledges the financial support from the French National Research Agency within the project ANR-24-EXMA-0001 PEPR MathsVivES CONDORCET.
  Kurata acknowledges support from KAKENHI (No.~25KJ2146) and Global Research Center (Waseda University).
  The authors' email addresses are:
  Federico Fioravanti, \protect\url{federico.fioravanti@univ-st-etienne.fr};
  Leo Kurata, \protect\url{leo.kurata.ac@gmail.com};
  Vasilis Stamatis, \protect\url{vs.stamatis@gmail.com}.
  During the preparation of this work, the authors used Claude (Anthropic) and ChatGPT (OpenAI) in order to assist with language editing and to identify possible algebraic or expository issues. After using these tools, the authors reviewed and edited the content as needed and take full responsibility for the content of the publication.}
\makeatother
\endgroup

\newpage

\newpage

\section{Introduction}\label{sec:introduction}

We study the setting where a set of individuals assigns each object to exactly one of the categories, using every category at least once, to obtain a collective classification. 
This problem, called classification aggregation, has been considered by \citet{maniquet2016theorem}, who show that an aggregator satisfying the properties of Unanimity, which requires society to respect agreement on an object's category, and Independence, which requires the collective assignment for that object to depend only on how voters classify it, must adopt the classification of one fixed voter (the dictator) at every profile.
In the preference aggregation setting \citep{arrow1963social}, where a finite society must turn any profile of rankings over at least three alternatives into a complete and transitive ranking of its own, unanimity and independence of irrelevant alternatives also force the existence of a dictator.
But non-dictatorial rules become possible when some assumptions are relaxed, such as the social ranking being quasi-transitive \citep{Gibbard2014} or being incomplete \citep{Weymark1984}.
In the same spirit, we study \emph{classification aggregation correspondences} (CACs).\footnote{See also \citet[Ch.~6]{Hervouin2025} for classification aggregation correspondences and their distinction from rules that return a set of single-valued classifications.}
Each voter still assigns one category to each object, but society may recognize several categories as final memberships of the same object, provided every object receives at least one category and every category remains represented.
Which non-dictatorial rules become possible under this relaxation?

Collective classification with multiple memberships can arise in a variety of settings. For example, consider a participatory-budgeting platform that asks residents to classify project proposals. Each resident chooses one primary category per proposal from a list that includes Arts, Biodiversity, and Climate.\footnote{The Pabulib data format illustrates projects with multiple thematic categories, such as Culture and Education or Health and Sport; see \url{https://pabulib.org/format}.}
Classifications help determine how public projects are compared and presented to citizens, and may also affect which sources of funding they can access.
An urban-greening project may aim both to restore habitats and to support climate adaptation.
One resident may therefore identify Biodiversity as its primary category, while another identifies Climate.
Listing the project under both headings recognizes both contributions. If eligibility for funding follows the assigned categories, the project could still be considered under Climate when the Biodiversity budget is exhausted.
It would retain both memberships even if only the Climate budget funded it.

In another application, a platform combines the predictions of several trained classifiers. Each classifier assigns one thematic label per project description, while the combined classification may retain several labels for the same project, a form of output also studied in multi-label learning \citep[e.g.,][]{gibaja2015tutorial}.

To isolate the effect of allowing multiple memberships, we admit every profile of individual classifications in the domain of \citet{maniquet2016theorem}.\footnote{We require each report to use every category to retain the individual domain of the benchmark model. The relevance of this restriction to participatory budgeting or learned classifications depends on the application.}
The requirement that every category remain represented collectively also has a practical interpretation.
Some participatory-budgeting schemes allocate separate budgets to thematic categories.\footnote{For example, France's 2018 ``Mon projet pour la plan\`ete'' initiative allocated separate budgets to Energy, Circular Economy and Biodiversity; see \url{https://jeparticipe.bourgognefranchecomte.fr/projet-planete-deposez-projet-de-budget-participatif-28-fevrier/}.}
When these budgets cannot be transferred between categories, leaving a category empty would leave its budget with no eligible proposal; this motivates requiring the collective classification to represent every category.

We also retain Independence, while strengthening Unanimity to Supportiveness, which requires every category collectively assigned to an object to have been proposed for that object by at least one individual.
Supportiveness implies Unanimity in our setting and is itself a consequence of Independence and Unanimity in Maniquet and Mongin's original setting \citep[Lemma~1]{maniquet2016theorem}.
In this finite-population benchmark with at least three categories, Independence and Supportiveness therefore describe exactly the same class as Independence and Unanimity, namely the dictatorial rules (\cref{prop:benchmark}).
In this sense, our approach preserves their individual domain and requirement that every category be represented, while changing the admissible form of the collective outcome.
The requirement that no category be left empty concerns the collective classification as a whole, while Independence requires considering each object separately.

On the other hand, once multiple memberships are allowed, even under Independence, Unanimity no longer delivers Supportiveness when there are three or more categories.
The clearest example is the classification counterpart of an oligarchy.
It assigns an object to the category on which the members of a coalition agree and, when they disagree, to every category.
This rule satisfies both Independence and Unanimity, and it is non-dictatorial once the coalition has two members (\cref{prop:oligarchy-resolution}).
Under the membership reading, disagreement within the coalition can lead society to assert that the object belongs to categories that nobody, inside or outside the coalition, has proposed.
When the designated voters disagree between Biodiversity and Climate, for example, the urban-greening proposal would also be listed under Arts.
Supportiveness excludes such a deicision.
In \cref{subsec:supportiveness}, we further justify this strengthening of Unanimity from both the interpretational and the technical perspectives.

Accepting every proposed category gives an independent and supportive rule that is non-dictatorial when there are at least two individuals.
We ask how decision power over the acceptance of assignments can be distributed while allowing some supported assignments to be rejected.
In particular, can a rule treat individuals equally and still reject some proposed memberships if each accepted membership must have a supporter whose report suffices on its own?
In the participatory-budgeting example, listing the urban-greening proposal under Climate expresses the panel's judgement that the project contributes to climate adaptation.
The platform may give designated panel members the authority to secure the inclusion of a thematic category.
Our Responsibility axiom requires that, for each accepted assignment, some supporting voter's complete classification be sufficient to sustain it, whatever the other voters report.
This requirement is stronger than the mere existence of a supporter.
Because thematic categories can overlap, the collective classification can retain both memberships when voters identify different primary categories.
Different voters can therefore sustain different memberships, including memberships of the same proposal, and share decision power over the collective classification.
For aggregation of learned classifications, the same requirement gives designated classifiers the authority to secure the inclusion of a label.

Our main results characterise the distribution of decision power over collective memberships.
Under Independence and Supportiveness, Responsibility and Rejection Stability each characterise the junta rules on finite populations, while Split Stability characterises the subclass in which the junta depends only on the object (\cref{thm:pair-representation,thm:general_junta,thm:frrc-characterisation}).\footnote{
The first two characterisations hold for $m\geq p\geq2$; the Split Stability characterisation requires $m>p\geq3$, where $m$ is the number of objects and $p$ the number of categories.
}
A junta rule fixes, for each object and category, a coalition -- which we call a \emph{junta} -- and assigns the object to the category exactly when at least one member of the junta does.
To ensure that every object receive a category and every category be used, the juntas for each object must have a common member, as must the juntas for each category (\cref{thm:junta-common-member}).
Under Category Neutrality, the junta depends only on the object; further adding Object Neutrality gives a single junta for all objects and categories.
The three axioms reach these characterisations from different directions.
Responsibility asks that every assigned category be secured by the complete report of one of its supporters, in the spirit of the expertise conditions of \citet{fioravanti2025expert};
Split Stability asks that, when support for a category is divided between it and a category previously unused at that object, at least one of the two is accepted exactly when the original was, while the object's other memberships remain unchanged; and
Rejection Stability asks that combining supporters' unchanged complete reports from two rejecting profiles should not secure the rejected assignment.
Every member of a junta can include a category on her own, and different members can include different categories at the same object; this is possible only because multiple memberships are allowed.
Note that when outcomes are required to be singletons, every junta shrinks to the same individual and the dictatorship of \citet{maniquet2016theorem} is recovered.

We also examine whether individual responsibility and equal treatment of individuals allow some supported assignments to be rejected.
Under Independence, Supportiveness and Responsibility, adding Anonymity singles out the full-population junta rule $J^N$ (\cref{cor:anon-finite}).
Thus, under the maintained axioms, equal treatment and individual responsibility require the acceptance of every supported assignment.
The rule is also neutral for all sizes of the problem, so that multiple memberships escape the non-existence of anonymous and neutral rules found by \citet{CaillouxEtAl2025}.

It turns out that the coalitional structure behind these results is a \emph{grill} \citep{choquet1947}.
Grills are related to the filters, familiar from aggregation theory.
The decisive coalitions in the theorem of Maniquet and Mongin form an ultrafilter, and an ultrafilter is exactly a family that is both a filter and a grill (\cref{fact:filter_grill}).
Much of the literature on Arrow's theorem weakens collective rationality while keeping the Pareto principle and (binary) independence, and it finds that the decisive coalitions retain the filter property.
Quasi-transitive social preferences lead to oligarchies \citep{Gibbard2014}, transitive but incomplete social preferences lead to oligarchies \citep{Weymark1984}, and acyclic social preferences lead to prefilters \citep{Brown1975,BossertCato2020}.
Under our axioms, relaxing single-valuedness in classification aggregation corresponds to keeping the other half of the ultrafilter.
Grills are the duals of filters (\cref{lem:duality}), and the coalitions that win under the junta rule with junta, say $M$, are exactly the blocking coalitions of the oligarchy with oligarchs $M$.
In the Arrovian literature, each member of a decisive group can block a strict social preference.
In our setting, each member of a junta can secure the inclusion of a category.
A rule that lets each member of a group block every category she does not propose leaves divided objects with no category at all, violating the requirement that every object receive a category.
In this sense, the passage from the theorem of Maniquet and Mongin to junta rules parallels the passage from Arrow's theorem to oligarchies, with the roles of filters and grills exchanged.
Refer to \cref{subsec:translating} and \cref{tab:dual} for technical details.

Under Independence, Supportiveness and Category Neutrality, our three new axioms are equivalent on finite populations when $m>p\geq3$.
By contrast, on infinite populations, Responsibility is strictly stronger than the two stability axioms under these conditions.
Arrow's theorem fails in infinite societies, because non-dictatorial rules are obtained whenever an ultrafilter is free \citep{fishburn1970arrow,KirmanSondermann1972}, and the same holds for the theorem of Maniquet and Mongin (\cref{cor:mm-possibility}).
With multiple memberships, Split Stability, or Rejection Stability together with Category Neutrality, characterises the limiting junta rules, a class that extends the object-dependent junta rules and whose winning coalitions form grills that need not be principal (\cref{thm:limiting}).
The rule that assigns an object to every category with infinitely many supporters belongs to this class and is defined explicitly.
Responsibility yields junta rules on every population by directly identifying the individuals whose classifications can sustain an assignment (\cref{cor:FRR-infinite}).

\subsection*{Related literature}
The axiomatic study of classification began with partitions, in which the categories carry no labels and their number is not fixed in advance.
\citet{Mirkin1975} was the first to analyse the aggregation of equivalence relations axiomatically, and \citet{FishburnRubinstein1986} show that independence and unanimity leave exactly the conjunctive rules, under which two objects are equivalent in the aggregate partition if and only if they are equivalent for every member of a fixed nonempty coalition (see also \citealp{DimitrovEtAl2012}).
These rules are oligarchies, and they are available because the intersection of equivalence relations is again an equivalence relation: agreement within a coalition always defines an admissible outcome.
With designated categories this is no longer the case.
\citet{maniquet2016theorem} extend to three or more categories a result that \citet{KasherRubinstein1997} obtained for two, showing that Independence and Unanimity force dictatorship when individual and collective classifications must leave no category empty, and \citet{DokowHolzman2010nonbinary} obtain the counterpart for classifications that place at most one object in each category.
Subsequent work retains single-valued collective classifications and varies the remaining requirements.
\citet{CaillouxEtAl2024} weaken Unanimity to citizen sovereignty;
\citet{Craven2023} shows that domain restrictions permitting majority-based aggregation have very limited scope in this setting, and identifies a hybrid structure in which such aggregation is possible exactly when each object is eligible for at most two categories;
\citet{fioravanti2025expert} examines forms of individual authority over objects or categories;
\citet{CaillouxEtAl2025} drop Independence and study anonymity and neutrality;
and \citet{Craven2024} considers ranked categories.
\citet{filmus2026classification} characterises all independent rules whose outcome is always surjective, without further assumptions on the objectwise aggregation functions, and derives from it a quantitative version of the dictatorship conclusion.
\citet[Ch.~6]{Hervouin2025} also considers classification aggregation correspondences.
Relative to this literature, we keep the individual domain of \citet{maniquet2016theorem}, Independence and the requirement that every category be used, and change only the form of the collective outcome.

Group identification \citep{KasherRubinstein1997} is the case in which the objects to be classified are the individuals themselves.
Because a group may be left empty there, rules that give individuals authority over their own membership remain available, and the literature characterises the liberal rule together with the consent rules, which balance self-determination against the opinions of others \citep{SametSchmeidler2003,SungDimitrov2005,Miller2008}.
\citet{ChoJu2017} determine memberships in three or more groups simultaneously and show that independence of irrelevant opinions, together with symmetry and non-degeneracy, leads to the liberal rule.
In these models possibility comes from allowing categories to remain empty; we keep the requirement that every category be used and obtain possibility from multiple memberships instead.
Several papers associate an object with several categories through degrees of membership.
\citet{ChoPark2018} allow fractional memberships in group identification and characterise weighted-average and fractional consent rules, and fuzzy versions of classification and group identification are studied by \citet{Alcantud2019}, \citet{FioravantiFuzzyGroup2022} and \citet{FIORAVANTI2025102422}, the last of which obtains weighted arithmetic means.
Our memberships are not graded.
An object belongs fully to each category assigned to it, and assigning it to one category does not weaken its membership in another; accordingly, the rules we obtain are coalitional rather than averaging.

Reading each statement ``$x$ belongs to $r$'' as a binary issue turns a CAC into a binary aggregator whose admissible collective evaluations form a strictly larger set than the admissible individual ones: an individual assigns each object exactly one category, whereas society assigns it at least one.
Our problem therefore belongs to the family in which society is held to a weaker consistency requirement than individuals.
In judgment aggregation, allowing incomplete collective judgments while retaining deductive closure yields oligarchies \citep{DietrichList2008}, and \citet{DokowHolzman2010abstention} show that the conditions on the domain that lead to dictatorship without abstentions lead to oligarchy once abstentions are allowed, with Gibbard's oligarchy theorem among the corollaries.
In these models society may take a position on fewer issues than individuals do, so agreement within a coalition remains an admissible outcome.
In ours, society may accept more membership statements than any individual does, disagreement is resolved by inclusion rather than abstention, and oligarchies give way to juntas.
Graph aggregation \citep{EndrissGrandi2017} provides a close formal analogue.
Its groundedness axiom requires every collective edge to belong to some individual graph, as Supportiveness does for memberships, and \citet{EndrissGrandi2017} show that, for graph properties that are contagious and implicative, unanimous, grounded and independent rules that preserve the property are oligarchic.
Applied to each membership statement, the quota rules of \citet{DietrichList2007quota} accept a category when it has enough supporters; a quota of one for every pair gives $J^N$, and pair-specific quotas give procedures such as $F^4$ in \cref{ex:running-quota}.
Finally, when individual and collective evaluations are drawn from the same set, \citet{DokowHolzman2010nonbinary} and \citet{SzegedyXu2015} study which domains of non-binary evaluations force supportive or unanimous aggregation to be dictatorial.
The setting of \citet{maniquet2016theorem} is such a domain, with objects as issues and categories as positions, and CACs fall outside this framework precisely because the collective outcome need not belong to the individual domain.

As recalled earlier, the ultrafilter approach identifies the decisive coalitions of an Arrovian rule with an ultrafilter \citep{KirmanSondermann1972,Hansson1976}, and weakenings of collective rationality correspond to weakenings of this structure, from filters and oligarchies to prefilters (see also \citealp{BossertSuzumura2009}).
\citet{CatoEtAl2026} develop this approach in an abstract framework in which an aggregation rule may remain silent.
Under a stability condition, the decisive coalitions form a filter, and when the set of signals is finite they characterise the rules that approve a signal exactly when some coalition in a given filter or ultrafilter agrees on it.
These rules are agreement-based.
Junta rules are inclusion-based, and their winning coalitions form grills \citep{choquet1947}, the duals of filters.
To our knowledge, grills have not been used explicitly as the decisive structure of aggregation rules, although they appear under another name: \citet{cato2013quasi} shows that, for transitive but possibly incomplete social preferences, the quasi-decisive coalitions form a quasi-ultrafilter, which by \cref{fact:grill_ultra} is exactly a grill.
In his setting these are the coalitions that can prevent society from ranking an alternative at least as high as one that all their members prefer to it; here they are the coalitions that can secure a membership.

\citet{fishburn1970arrow} and \citet{KirmanSondermann1972} show that Arrow's conditions can be met by non-dictatorial rules on infinite populations, through free ultrafilters that cannot be explicitly constructed.
\citet{mihara1997arrow} shows that in a countably infinite society every rule satisfying Arrow's conditions that is pairwise computable is dictatorial, and \citet{mihara1997anonymity} shows that in infinite societies anonymity is inconsistent with Arrow's unanimity and independence conditions on sufficiently rich domains.
\citet{torres2005limiting} introduces limiting dictatorial rules, of which our limiting junta rules are the counterparts, and \citet{Cato2021,Cato2022} study further aspects of preference aggregation with infinite populations.
In group identification, \citet{FioravantiTohme2020} show that, with infinitely many voters, the characterisation of the liberal rule survives whereas that of the oligarchic rule does not.
A similar divergence appears here.
The axiom that identifies individual supporters, Responsibility, characterises junta rules on every population, whereas the axioms that only relate coalitions, Split Stability and Rejection Stability, admit non-principal grills.
The resulting rules include the Fr\'echet rule, which, unlike the rules built on free ultrafilters, is defined without the axiom of choice.

\subsection*{Paper organisation}
The paper is organized as follows.
\Cref{sec:preliminaries} introduces the model and the coalitional families we use.
\Cref{sec:axioms} presents the basic axioms and the benchmark.
\Cref{sec:oligarchy_rules} discusses the two translations of oligarchy and introduces junta rules.
\Cref{sec:characterisation} contains the three characterisations, examines the role of Supportiveness, and concludes by relating Anonymity to responsibility.
\Cref{sec:infinite} turns to infinite populations.
\Cref{sec:conclusion} concludes.
Longer proofs are collected in \cref{app:proofs}, and \cref{app:independence} shows that the axioms in the main characterisations are independent.
\section{Preliminaries}\label{sec:preliminaries}

\subsection{Notation}\label{sec:notation}

Let $N$ be a set of individuals with at least two members.
Unless stated otherwise, $N=\{1,\ldots,n\}$ is finite; in \cref{sec:infinite} and wherever we say so explicitly, $N$ may be infinite, for instance $N=\N=\{1,2,\ldots\}$.
Let $X=\{x_1,\ldots,x_m\}$ be a finite set of $m$ objects and $P$ a finite set of $p$ categories, with $m\geq p\geq 2$.

An \emph{individual classification} is a surjective mapping $c\colon X\to P$.
We denote the set of individual classifications by $\C=\{c\colon X\to P\mid c\text{ is surjective}\}$.
A \emph{social classification} is a mapping $\bar{c}\colon X\to 2^{P}\setminus\{\emptyset\}$ satisfying $\bigcup_{x\in X}\bar{c}(x)=P$.
We denote the set of social classifications by $\Cbar$.

A \emph{profile} is a vector $\cl=(c_i)_{i\in N}\in\C^N$ of individual classifications.
A \emph{classification aggregation correspondence} (CAC) is a mapping $F\colon\C^{N}\to\Cbar$ that assigns a social classification to every profile.
A CAC is \emph{resolute} if $F(\cl)(x)$ is a singleton for all $\cl$ and $x$; resolute CACs are the classification aggregation \emph{functions} studied by \citet{maniquet2016theorem}.
The rule they specify is a \emph{dictatorial} rule, defined by the existence of $d\in N$ such that $F(\cl)(x)=\{c_d(x)\}$ for all $\cl\in\C^N$ and $x\in X$.

Given a profile $\cl$ and an object $x\in X$, we write
\begin{align}
  S_{r}(x,\cl)=\{i\in N \mid c_{i}(x)=r\}
\end{align}
for the \emph{support set} of category $r$ at object $x$.
Given $\cl\in\C^N$, we write $\cl_x=\bigl(c_i(x)\bigr)_{i\in N}\in P^N$ for the \emph{opinion vector} at~$x$.
Given an opinion vector $\pi\in P^N$, $S_r(\pi)=\{i\in N\mid\pi(i)=r\}$ denotes the support set of category $r$.

Any opinion vector at a fixed object can be obtained from a profile of individual classifications.
Note that both lemmas below hold for finite and infinite populations.

\begin{lemma}\label{lem:real}
  Assume $m\geq p \ge 2$.
  For every $\pi\in P^{N}$ and every $x\in X$, there exists a profile $\cl\in\C^{N}$ such that $\cl_x=\pi$.
\end{lemma}

\begin{proof}
  Fix $\pi\in P^N$ and $x\in X$.
  For each $i\in N$, set $c_i(x)=\pi(i)$.
  It remains to define $c_i$ on $X\setminus\{x\}$ so that $c_i$ is surjective.
  The category $\pi(i)$ is already attained at~$x$.
  Since $|X\setminus\{x\}|=m-1\geq p-1=|P\setminus\{\pi(i)\}|$, we can assign every category in $P\setminus\{\pi(i)\}$ to at least one object in $X\setminus\{x\}$ and assign any remaining objects arbitrarily.
\end{proof}

When $m>p$, the profile can be chosen so that changing the opinion vector at $x$ never affects surjectivity.

\begin{lemma}\label{lem:rich}
  Assume $m > p \ge 2$.
  For every $\pi\in P^{N}$ and every $x\in X$, there exists a profile $\cl\in\C^N$ such that $\cl_x=\pi$ and, for every $i\in N$, the restriction $c_{i}|_{X\setminus\{x\}}$ is surjective.
\end{lemma}

\begin{proof}
  Fix $\pi\in P^{N}$ and $x\in X$, and write $P=\{r_1,\dots,r_p\}$.
  For each $i\in N$, set $c_{i}(x)=\pi(i)$.
  Since $|X\setminus\{x\}|=m-1\geq p$, we may choose pairwise distinct objects $y_{1},\dots,y_{p}\in X\setminus\{x\}$ and set $c_{i}(y_{j})=r_j$ for $j\in\{1,\dots,p\}$, and $c_{i}(y)=\pi(i)$ for every remaining $y\in X\setminus\{x\}$.
  Then $c_{i}|_{X\setminus\{x\}}$ is surjective, hence so is $c_{i}$.
\end{proof}

A set-valued outcome of CACs admits two readings: membership or tie.
Under the \emph{membership reading}, $r\in F(\cl)(x)$ means that society assigns $x$ to $r$, so that $x$ may belong to several categories at once.
Under the \emph{tie reading}, $F(\cl)(x)$ lists the categories that remain admissible for $x$, and a single category is to be selected later.
We adopt the membership reading throughout.
The covering condition $\bigcup_{x\in X}\bar c(x)=P$ in the definition of a social classification is formulated for this reading.
It says that every category has at least one member, which is how the surjectivity requirement translates when an object may belong to several categories.
Under the tie reading, we would instead require the existence of a surjective selection: a surjective mapping $c\colon X\to P$ with $c(x)\in\bar c(x)$ for every $x\in X$.
Under this reading a single category must eventually be selected for each object, and if this selection respects the two postulates of \citet{maniquet2016theorem}, their impossibility applies at that final stage.
Under the membership reading, the set-valued outcome is the final classification.

\subsection{Examples of CACs}\label{sec:example}

We introduce four running examples involving five rules at a common profile $\cl$.
The examples distinguish whether accepted assignments have supporters, whether an object's classification depends on reports about other objects, and whether an individual's report can secure an assignment.
They also compare individual decision power with equal treatment of voters.
We return to these examples when introducing the axioms and summarise their properties in \cref{tab:examples}.

A participatory-budgeting platform groups project proposals into thematic categories for comparison.
Two voters, $N=\{1,2\}$, classify four proposals, $X=\{x_1,x_2,x_3,x_4\}$, into three categories, $P=\{a,b,c\}$, which we read as Arts, Biodiversity and Climate, respectively.
Each voter assigns every proposal to one category and uses every category, whereas the platform may place a proposal in several categories, for instance a project combining public art and habitat restoration in Arts and in Biodiversity.
The list $\bigl(c(x_1),c(x_2),c(x_3),c(x_4)\bigr)\in P^X$ denotes the classification $c$, and a social classification $\bar c$ is written in the same way.
In each of the examples below, the profile $\cl=(c_1,c_2)$ is given by
\begin{align}
  c_1&=(a,b,c,a),\\
  c_2&=(a,a,b,c).
\end{align}
The voters agree on $x_1$ and disagree on the other three proposals.
Observe that both classifications are surjective.

\begin{example}\label{ex:running-oligarchy}
  Let $F^1(\cl)(x)=\{c_1(x)\}$ if $c_1(x)=c_2(x)$, and $F^1(\cl)(x)=P$ otherwise.
  At the profile above,
  \begin{align}
    F^1(\cl)=(\{a\},P,P,P).
  \end{align}
  Every category is used, so $F^1(\cl)$ is a social classification, and the same holds at every profile.
  The outcome for a proposal depends only on how the two voters classify that proposal, and a proposal on which they agree is placed exactly in the category they chose.
  At $x_2$, however, voter~1 proposes Biodiversity and voter~2 Arts, and $F^1$ places the proposal in Climate as well, although neither of them proposed it.
  In this case, the rule uses every category only because it places each disputed proposal in all of them.  \hfill $\diamond$
\end{example}

\begin{example}\label{ex:running-junta}
  Let $F^2(\cl)(x)=\{c_1(x),c_2(x)\}$, so that each proposal is placed in every category that some voter proposes for it.
  At the profile above,
  \begin{align}
    F^2(\cl)=(\{a\},\{a,b\},\{b,c\},\{a,c\}).
  \end{align}
  Like $F^1$, this rule depends only on the assignments for each proposal and respects agreement, but it never places a proposal in a category that nobody proposed.
  The voters whose assignments count may also vary across proposals and categories.
  Let $F^{2'}$ coincide with $F^2$, except that Biodiversity is assigned to $x_2$ only when voter~2 proposes it, and Climate is assigned to $x_4$ only when voter~1 proposes it, for instance because voter~2 has expertise in biodiversity conservation and voter~1 in climate adaptation.
  At the profile above,
  \begin{align}
    F^{2'}(\cl)=(\{a\},\{a\},\{b,c\},\{a\}).
  \end{align}
  Every category is still used, and $F^{2'}$ produces a social classification at every profile  (see also the discussion in \cref{ex:running-quota}). \hfill $\diamond$
\end{example}

Note that, under the tie reading of set-valued outcomes, no selection from $F^{2'}(\cl)$ in \cref{ex:running-junta} is surjective, whereas the membership reading is still applicable.

\begin{example}\label{ex:running-split}
  Let $F^3(\cl)(x)$ consist of $c_1(x)$ together with $c_2(x)$ whenever voter~2 uses $c_2(x)$ for at least one other proposal.
  Voter~2's assignment for a proposal thus counts only if she uses that category more than once; at the profile above, this is the case only for Arts.
  Since $F^3(\cl)$ always contains voter~1's classification, it is a social classification at every profile, and every category it assigns is proposed by someone.
  At the profile above,
  \begin{align}
    F^3(\cl)=(\{a\},\{a,b\},\{c\},\{a\}).
  \end{align}
  The rule also uses every category. \hfill $\diamond$
\end{example}

Lastly, the platform may instead use a \emph{quota rule}, requiring a specified minimum number of supporting voters for each proposal--category pair.
If one supporter suffices for every pair, this procedure coincides with $F^2$.
With two voters, a quota of~$2$ requires both of them to propose the category.
Raising every quota to~$2$ gives $(\{a\},\emptyset,\emptyset,\emptyset)$ at the profile above: three proposals are left without a category, and Biodiversity and Climate are not used.

\begin{example}\label{ex:running-quota}
  Let $F^4$ require two supporters for Art at $x_1$, for Biodiversity at $x_2$, and for Climate at $x_4$, and one supporter for every other pair.
  At the profile above, this rule yields
  \begin{align}
    F^4(\cl)=(\{a\},\{a\},\{b,c\},\{a\}).
  \end{align}
  This rule produces a social classification at every profile.
  Its outcome at the profile above coincides with that of $F^{2'}$ in \cref{ex:running-junta}, but the reasons behind them are different.
  Under $F^{2'}$, voter~1's assignment of Biodiversity to $x_2$ does not count; under $F^4$, it counts but is not enough on its own.
  If the voters exchange their reports, so that voter~1 submits $c_2$ and voter~2 submits $c_1$, $F^{2'}$ gives $(\{a\},\{a,b\},\{b,c\},\{a,c\})$, whereas $F^4$ gives the same outcome as before, since the quotas count supporters without regard to who they are.
  Yet neither voter can secure Biodiversity for $x_2$ by herself. In these cases, all the categories are used because all quotas are reached (which might not always be the case). \hfill $\diamond$
\end{example}

\subsection{Filters, ultrafilters and grills}\label{sec:filters}

We use families of coalitions to describe which groups can secure an assignment.
Filters and related families also describe decisive coalitions in preference aggregation \citep{sen1970collective,sen1995rationality}.
We recall the definitions used below.

\begin{definition}\label{def:families}
  A family $\W$ of subsets of $N$ may have the following properties:
  \begin{enumerate}[label=(p.\arabic*),noitemsep]
    \item\label{p:emptyset} $\emptyset\notin\W$.
    \item\label{p:wholeset} $N\in\W$.
    \item\label{p:upward} For all $A,B\subseteq N$, if $A\in\W$ and $A\subseteq B$, then $B\in\W$.
    \item\label{p:filter} For all $A,B\in\W$, $A \cap B\in\W$.
    \item\label{p:ultra} For all $A\subseteq N$, $A\notin\W$ implies $N\setminus A\in\W$.
    \item\label{p:grill} For all $A,B\subseteq N$, $A \cup B\in\W$ implies $A\in\W$ or $B\in\W$.
  \end{enumerate}
  A family $\W$ of subsets of $N$ is called
  \begin{itemize}[noitemsep,leftmargin=3.5em]
    \item a \emph{filter}, denoted by $\F$, if it satisfies \ref{p:emptyset}--\ref{p:filter};
    \item an \emph{ultrafilter}, denoted by $\U$, if it satisfies \ref{p:emptyset}--\ref{p:ultra}; and
    \item a \emph{grill}, denoted by $\G$, if it satisfies \ref{p:emptyset}--\ref{p:upward} and \ref{p:grill}.
  \end{itemize}
\end{definition}

Property \ref{p:grill} is called the \emph{grill property}.
Equivalently, letting $\I=2^{N}\setminus\G$, the grill property states that $\I$ is closed under union: if $A\notin\G$ and $B\notin\G$, then $A\cup B\notin\G$.
By induction, if $A_1\cup\cdots\cup A_k\in\G$, then $A_j\in\G$ for some~$j$.

A filter is \emph{principal} if there is a nonempty coalition $S \in \F$ such that $T \supseteq S$ for all $T \in\F$.
In that case $\F=\F_S$, where $\F_S=\{T\subseteq N\mid T\supseteq S\}$.
An ultrafilter is principal if and only if it equals $\F_{\{i\}}$ for some $i\in N$, and it is \emph{free} otherwise.
A grill $\G$ on $N$ is \emph{principal} if $\G =\G_M$ for some nonempty $M \subseteq N$, where
$$
\G_M=\{A \subseteq N \mid A \cap M \neq \emptyset\},
$$
and it is \emph{free} otherwise.
Letting $M_{\G} = \{i \in N \mid \{i\} \in \G\}$, every coalition meeting $M_{\G}$ belongs to $\G$ by \ref{p:upward}.
Hence, $\G$ is principal if and only if every member of $\G$ meets $M_{\G}$, in which case $\G=\G_{M_\G}$.
When $N$ is finite, all filters, ultrafilters and grills on $N$ are principal; for grills, this is the content of the next fact.

\begin{fact}\label{prop:grill-structure}
  Let $N$ be finite.
  A family $\G\subseteq2^N$ is a grill on $N$ if and only if there exists a unique nonempty $M\subseteq N$ such that $\G=\G_M$.
\end{fact}

\begin{proof}
  Let $\G$ be a grill and $\I=2^{N}\setminus\G$.
  By the contrapositive of \ref{p:upward}, $\I$ is downward closed, and by the contrapositive of \ref{p:grill}, it is closed under union.
  Since $\emptyset\in\I$ and $N$ is finite, $S=\bigcup_{A\in\I} A$ belongs to $\I$ and $\I=2^{S}$.
  Let $M=N\setminus S$.
  Then $M\neq\emptyset$ because $N\in\G$, and $A\in\G$ if and only if $A\not\subseteq S$, that is, if and only if $A\cap M\neq\emptyset$.
  Uniqueness holds because $M=M_\G$.
  Conversely, $\G_M$ satisfies \ref{p:emptyset}--\ref{p:upward} when $M\neq\emptyset$, and \ref{p:grill} holds because $(A\cup B)\cap M=(A\cap M)\cup(B\cap M)$.
\end{proof}

We record three further properties of grills.

\begin{fact}\label{fact:grill_ultra}
  Let $\W$ be a family of subsets of $N$ satisfying \ref{p:emptyset}--\ref{p:upward}.
  If $\W$ satisfies \ref{p:grill}, then it satisfies \ref{p:ultra}.
\end{fact}

\begin{proof}
  Take any $A\subseteq N$.
  Since $A\cup(N\setminus A)=N\in\W$ by \ref{p:wholeset}, property \ref{p:grill} gives $A\in\W$ or $N\setminus A\in\W$.
\end{proof}

\citet{cato2013quasi} introduces the notion of a quasi-ultrafilter, defined as a family satisfying \ref{p:emptyset}--\ref{p:upward}, \ref{p:ultra}, and \ref{p:grill}.
By \cref{fact:grill_ultra}, condition \ref{p:ultra} is redundant in this definition, so a quasi-ultrafilter is precisely a grill.
\Cref{fact:grill_ultra} also leads to the following characterisation of ultrafilters.

\begin{fact}\label{fact:filter_grill}
  A family of subsets of $N$ is both a filter and a grill if and only if it is an ultrafilter.
\end{fact}

\begin{proof}
  Let $\U$ be an ultrafilter.
  It is a filter by definition, so it remains to verify \ref{p:grill}.
  Suppose, to the contrary, that $A\cup B\in\U$ but $A\notin\U$ and $B\notin\U$ for some $A,B\subseteq N$.
  By \ref{p:ultra}, $N\setminus A\in\U$ and $N\setminus B\in\U$, and so \ref{p:filter} gives $N\setminus(A\cup B)\in\U$.
  Another application of \ref{p:filter} yields $\emptyset=(A\cup B)\cap\bigl(N\setminus(A\cup B)\bigr)\in\U$, which contradicts \ref{p:emptyset}.

  Conversely, let $\W$ be both a filter and a grill.
  Then $\W$ satisfies \ref{p:emptyset}--\ref{p:filter}, and it satisfies \ref{p:ultra} by \cref{fact:grill_ultra}.
  Hence, $\W$ is an ultrafilter.
\end{proof}

For instance, consider $|N|\geq 2$ and $\W=\{A\subseteq N : A\neq\emptyset\}$.
This is a grill but not an ultrafilter because $\{i\}\cap\{j\}=\emptyset$ for any distinct $i,j\in N$.

In fact, grills and filters are dual to each other \citep{choquet1947}.
For a family $\G\subseteq 2^N$, let $\G^{\ast}=\{N\setminus A\mid A\subseteq N,\ A\notin\G\}$ be the family of complements of the sets outside $\G$.
We call the members of $\G^\ast$ the \emph{blocking coalitions} of $\G$.

\begin{fact}\label{lem:duality}
  Let $\G\subseteq 2^N$.
  \begin{enumerate}[label=(\roman*),noitemsep]
    \item $\G$ is a grill if and only if $\G^\ast$ is a filter.
    \item If $\G$ is a grill, then $S\in\G$ if and only if $S\cap T\neq\emptyset$ for every $T\in\G^\ast$.
    \item For every nonempty $M\subseteq N$, $(\G_M)^\ast=\F_M$ and $(\F_M)^\ast=\G_M$. In particular, a grill is principal if and only if its family of blocking coalitions is a principal filter.
  \end{enumerate}
\end{fact}

\begin{proof}
  (i) The map $A\mapsto N\setminus A$ is a bijection from $2^N\setminus\G$ onto $\G^\ast$.
  Under it, \ref{p:emptyset} for $\G$ corresponds to \ref{p:wholeset} for $\G^\ast$, \ref{p:wholeset} for $\G$ to \ref{p:emptyset} for $\G^\ast$, and upward closure of $\G$ to upward closure of $\G^\ast$.
  Finally, the grill property of $\G$ says that $A,B\notin\G$ implies $A\cup B\notin\G$, which is equivalent to $(N\setminus A)\cap(N\setminus B)\in\G^\ast$ whenever $N\setminus A,N\setminus B\in\G^\ast$, that is, to \ref{p:filter} for $\G^\ast$.

  (ii) If $S\in\G$ and $S\cap T=\emptyset$ for some $T\in\G^\ast$, then $S\subseteq N\setminus T\notin\G$, contradicting upward closure.
  If $S\notin\G$, then $T=N\setminus S\in\G^\ast$ and $S\cap T=\emptyset$.

  (iii) $A\notin\G_M$ if and only if $A\cap M=\emptyset$, that is, if and only if $M\subseteq N\setminus A$.
  Likewise, $B\in(\F_M)^\ast$ if and only if $M\not\subseteq N\setminus B$ if and only if $B\cap M\neq\emptyset$.
\end{proof}

\section{Basic axioms and the Maniquet--Mongin benchmark}\label{sec:axioms}

\subsection{Independence and Unanimity}\label{subsec:axioms}

The first axiom states that the social classification of an object depends only on how the individuals classify that object.

\begin{axiom}[Independence]
  \label{ax:I}
  A CAC $F$ satisfies \emph{Independence} if, for all $x\in X$ and all $\cl,\cl'\in\C^{N}$, $c_{i}(x)=c'_{i}(x)$ for every $i\in N$ implies $F(\cl)(x)=F(\cl')(x)$.
\end{axiom}

Under Independence, $F$ is determined by the family of mappings $(f_x)_{x\in X}$ with $f_x\colon P^N\to2^P\setminus\{\emptyset\}$ and $f_x(\cl_x)=F(\cl)(x)$; by \cref{lem:real}, every $f_x$ is defined on all of $P^N$.

\citet{maniquet2016theorem} use the following requirement together with Independence.

\begin{axiom}[Unanimity]
  \label{ax:U}
  A CAC $F$ satisfies \emph{Unanimity} if, for all $x\in X$ and all $\cl\in\C^N$ such that $c_i(x)=c_j(x)$ for every $i,j\in N$, it is the case that $F(\cl)(x)=\{c_i(x)\}$.
\end{axiom}

These are the two requirements in \citet{maniquet2016theorem}.
Allowing multiple memberships makes it necessary to examine whether Unanimity alone provides an adequate connection between individual reports and collectively accepted assignments.

\subsection{Supportiveness beyond unanimity}\label{subsec:supportiveness}

Unanimity constrains the outcome when all individuals agree, but it does not in general prevent an unsupported assignment when they disagree.
Because our setting allows multimembership, we propose the use of Supportiveness, which strengthens Unanimity.
Supportiveness requires every socially assigned category to have at least one individual supporter.

\begin{axiom}[Supportiveness]
  \label{ax:S}
  A CAC $F$ satisfies \emph{Supportiveness} if, for all $x\in X$ and all $\cl\in\C^{N}$,
  $$
  F(\cl)(x)\subseteq\{c_{i}(x) \mid i\in N\}.
  $$
\end{axiom}

Supportiveness can be read as the unanimity principle applied to each membership statement ``$x$ belongs to $r$''.
If every individual rejects the statement, so does society.
If every individual accepts it, the outcome at $x$ is a nonempty subset of $\{r\}$, so society accepts it as well.
The requirement that society reject unsupported memberships has several precedents.
In the single-valued setting, \citet[Lemma~1]{maniquet2016theorem} derive it from their axioms, and it is part~(b) of the unary unanimity condition of \citet{Craven2023}.
It is the supportiveness condition of \citet{DokowHolzman2010nonbinary} for non-binary evaluations, it corresponds to the zero unanimity of \citet{FIORAVANTI2025102422} for fuzzy classification aggregation, the groundedness axiom of \citet{EndrissGrandi2017} for graph aggregation, and the Pareto$^{\!\star\!}$ principle of \citet{kurata2025reservation, kurata2026risk} for Harsanyian aggregation of incomplete preferences.

Supportiveness implies Unanimity in our framework.
For let $F$ be a CAC and suppose that every individual assigns $x$ to $r$; by Supportiveness, $F(\cl)(x)\subseteq\{r\}$, and $F(\cl)(x)\neq\emptyset$ follows from $F(\cl)\in\Cbar$, meaning $F(\cl)(x)=\{r\}$.
The converse holds only when $p=2$ and fails when $p\geq3$.

\begin{lemma}\label{lem:S-implies-U}
  Every CAC $F$ that satisfies Supportiveness satisfies Unanimity.
\end{lemma}

Among the rules of \cref{ex:running-oligarchy,ex:running-junta,ex:running-split,ex:running-quota}, only $F^3$ violates Independence: at $x_2$ and $x_4$ its outcome changes although nobody changes her classification of these proposals.
Only $F^1$ violates Supportiveness: at $x_2$ it adds Climate, which neither voter proposes.

\Cref{sec:oligarchy_rules,subsec:role-supp} show why Unanimity cannot replace Supportiveness in the Responsibility characterisation when $p\geq3$.
Although the two axioms are interchangeable in some of our results, Unanimity alone is insufficient under the membership reading.
We adopt Supportiveness throughout so that every accepted assignment has an individual supporter.
The two axioms differ when individuals disagree.
Suppose that some voters classify a proposal as Arts and the others as Biodiversity.
Listing the proposal under both categories is a reasonable way of reflecting this disagreement, since each category records the judgement of some members and can be defended by pointing to them.
Listing it under Climate as well is different in kind.
Nobody proposed Climate for the proposal, so no voter can be asked to justify that assignment, and a resident comparing proposals in Climate would find one that no voter placed there.
Unanimity is silent about this profile, since the members do not agree, and so it does not prevent Climate from being added.
Supportiveness excludes Climate and allows Arts, Biodiversity or both to be accepted.
The rule $F^1$ of \cref{ex:running-oligarchy} behaves in exactly this way at $x_2$: it satisfies Unanimity, yet it lists the proposal under Climate, whereas $F^2$ lists it under Arts and Biodiversity only.

For resolute rules, the distinction does not arise, because there Unanimity, Independence and the requirement that every category be used already imply Supportiveness.

\begin{proposition}\label{prop:benchmark}
  Let $m\geq p\geq 2$ and $m\geq3$.
  For a resolute CAC $F$, the following statements are equivalent.
  \begin{enumerate}[label=(\roman*),noitemsep]
    \item $F$ satisfies Independence and Unanimity.
    \item $F$ satisfies Independence and Supportiveness.
    \item $F$ is dictatorial.
  \end{enumerate}
\end{proposition}

\begin{proof}
  A dictatorial rule satisfies Independence and Supportiveness, so (iii) implies (ii).
  By \cref{lem:S-implies-U}, (ii) implies (i).
  That (i) implies (iii) is the theorem of \citet{maniquet2016theorem} when $p\geq3$ and Corollary~2 of \citet{CaillouxEtAl2024} when $p=2$.
\end{proof}

\Cref{prop:benchmark} shows that replacing Unanimity with Supportiveness preserves the benchmark characterisation of \citet{maniquet2016theorem}.
In their setting it is already implied by Independence, Unanimity and the requirement that the social classification be surjective.
In our setting the requirement on the social classification is weaker, because a category can be used by assigning objects to several categories at once, and this derivation of Supportiveness is no longer available.
Independence and Unanimity therefore do not suffice to exclude unsupported memberships.

\subsection{Neutrality}

We also consider symmetry requirements on objects and categories.
The next axiom requires that the names of the objects be irrelevant for the aggregation.

\begin{axiom}[Object Neutrality]
  \label{ax:ON}
  A CAC $F$ satisfies \emph{Object Neutrality} if, for all $\cl\in\C^{N}$, all permutations $\mu\colon X\to X$ and all $x\in X$, $F(\cl\circ\mu)(x)=F(\cl)(\mu(x))$, where $\cl\circ\mu=(c_i\circ\mu)_{i\in N}$.
\end{axiom}

If $\cl\in\C^N$ and $\mu$ is a permutation of $X$, then $\cl\circ\mu\in\C^N$, since the composition of a surjection with a bijection is a surjection.
Category Neutrality requires that the aggregation rule treat all categories symmetrically.

\begin{axiom}[Category Neutrality]
  \label{ax:CN}
  A CAC $F$ satisfies \emph{Category Neutrality} if, for every permutation $\sigma$ of $P$ and every $\cl\in\C^{N}$, $F(\sigma\circ\cl)(x)=\sigma(F(\cl)(x))$ for all $x\in X$, where $\sigma\circ\cl=(\sigma\circ c_i)_{i\in N}$ and $\sigma(A)=\{\sigma(r) \mid r\in A\}$ for $A\subseteq P$.
\end{axiom}

Notice that in the running examples, all rules satisfy both neutrality axioms except for $F^{2'}$ and $F^4$: the former makes whose assignment counts depend on the proposal and category, and the latter makes the required quota depend on them.

\section{Oligarchy and junta rules}\label{sec:oligarchy_rules}

\subsection{Translating oligarchy into a CAC}\label{subsec:translating}

In preference aggregation, an oligarchy is a coalition $M$ whose unanimous strict preference determines the social strict preference, while opposing strict preferences among its members result in social indifference \citep{Gibbard2014}.
There are two ways to carry this idea over to classification: the agreement-based and inclusion-based translations.

The \emph{agreement-based} translation retains a category only when every member of $M$ assigns $x$ to it, that is, it selects $\{r\in P\mid M\subseteq S_r(x,\cl)\}=\bigcap_{i\in M}\{c_i(x)\}$.
This set becomes empty as soon as two members of $M$ classify $x$ differently, so it defines a CAC only when $M$ is a singleton.\footnote{Allowing an empty outcome would amount to suspending judgement, as in the weak oligarchies of \citet{DietrichList2008}. A CAC, however, has to classify every object.}
An agreement-based CAC must therefore specify what happens when the members of $M$ disagree.
We consider that the counterpart of social indifference in our setting is to assign $x$ to every category.
For a nonempty coalition $M$, the \emph{\textbf{oligarchy rule (with all-category assignment)}} is defined by
\begin{align}\label{eq:O-rule}
  O^M_{+}(\cl)(x)
  =
  \begin{cases}
    \{r\} &\text{if } M\subseteq S_{r}(x,\cl),\\
    P &\text{otherwise.}
  \end{cases}
\end{align}

The \emph{inclusion-based} translation retains every category to which at least one member of $M$ assigns $x$:
\begin{align}\label{eq:J-rule}
  J^M(\cl)(x)
  =
  \left\{ r\in P \mid S_{r}(x,\cl)\cap M\neq\emptyset\right\}
  =
  \{c_i(x) \mid i\in M\}.
\end{align}
We call $M$ a \emph{junta} and $J^M$ the \emph{\textbf{uniform junta rule}}.
No restriction is imposed on the size of $M$.
With $M=N$, the rule $O^N_{+}$ is $F^1$ of \cref{ex:running-oligarchy}, and $J^N$ is $F^2$ of \cref{ex:running-junta}.
At the profile of these examples, the two rules agree at $x_1$, where the voters agree, and differ at the three disputed proposals, where $F^1$ returns $P$ and $F^2$ returns only the proposed categories.

Both rules give the members of $M$ the same power over exclusive assignments: under either rule, the outcome at $x$ is $\{r\}$ if and only if $S_r(x,\cl)\in\F_M$.
They differ in how categories are included.
Under $J^M$, category $r$ is included if and only if $S_r(x,\cl)\in\G_M$, whereas under $O^M_{+}$ whether $r$ is included is not determined by the support of $r$ alone, since disagreement among the other members of $M$ puts every category into the outcome.
By \cref{lem:duality}, $\G_M$ consists exactly of the sets whose complements are not in $\F_M$.

\begin{proposition}\label{prop:oligarchy-resolution}
  Let $m\geq p\geq 2$ and let $M\subseteq N$ be nonempty.
  \begin{enumerate}[label=(\roman*),noitemsep]
    \item $O^M_{+}$ is a CAC that satisfies Independence, Unanimity, Category Neutrality and Object Neutrality.

    \item $O^M_{+}$ is dictatorial if and only if $|M|=1$.
    \item If $p=2$, then $O^M_{+}=J^M$.
    \item If $p\geq3$ and $|M|\geq2$, then $O^M_{+}$ violates Supportiveness.
  \end{enumerate}
\end{proposition}

The proof is in \cref{app:proofs}.
It follows from (i) and (iii) of \cref{prop:oligarchy-resolution} and \cref{lem:S-implies-U} that when $m\geq p\geq 2$, every CAC that satisfies Unanimity satisfies Supportiveness if and only if $p=2$.

The oligarchy rule is thus the most direct translation of an oligarchy, and it satisfies the axioms of \citet{maniquet2016theorem} without being dictatorial.
We nevertheless regard it as unsatisfactory.
Under the membership reading of \cref{sec:notation}, the outcome $P$ asserts that the object belongs to every category, including categories to which nobody, inside or outside $M$, has assigned it.
In the participatory-budgeting example, a proposal on which two members of the oligarchy disagree would be placed in every category, as $x_2$, $x_3$ and $x_4$ are under $F^1$ in \cref{ex:running-oligarchy}.
The rule also satisfies the requirement that every category be used in an uninformative way: whenever two oligarchs disagree at some object, that object alone covers all categories.
The inclusion-based translation satisfies Supportiveness.
Under the tie reading, by contrast, the outcome $P$ would mean that the question is left open, and $O^M_{+}$ would be a reasonable counterpart of an oligarchy; the choice between the two readings is therefore substantive.

The two translations also differ in their coalitional structure.
Under $J^M$, the coalitions that secure the inclusion of a category form the grill $\G_M$, which satisfies \ref{p:grill} but is not closed under intersection when $|M|\geq2$, and each member of $M$ can include a category on her own.
Since $\G_M$ consists of the sets whose complements are not in $\F_M$, these are exactly the coalitions that can prevent a strict social preference under the oligarchy with oligarchs $M$: a coalition that can block a strict preference there can force an assignment here.
Agreement, by contrast, cannot serve as the criterion for inclusion.
If the coalitions that can secure the inclusion of a category formed a filter $\F_M$ with $|M|\geq2$, every member of $M$ could veto the inclusion of each category she does not propose, which is the counterpart of an oligarch's veto,\footnote{Formally, an individual $i\in N$ has \emph{veto power} under a CAC $F$ if, for all $\cl\in\C^{N}$, all $x\in X$ and all $k\in P$, $c_{i}(x)\neq k$ implies $k\notin F(\cl)(x)$.}
and a profile at which two members of $M$ disagree would leave the object without any category.
Such a filter must therefore be an ultrafilter, which on a finite population means a dictatorship.
Indeed, an individual with veto power is a dictator, since the outcome at every object is then a nonempty subset of $\{c_i(x)\}$.

\subsection{The class of junta rules}\label{sec:junta}

The uniform junta rule is a special case of the general class in which the junta may depend on both the object and the category.
For a family of coalitions $(M_{x,r})_{x\in X,r\in P}$, the \emph{\textbf{general junta rule}} is defined, for all $\cl\in\C^N$ and all $x\in X$, by
\begin{equation}\label{eq:pair-J-rule}
  J^{(M_{x,r})}(\cl)(x)=\{r\in P\mid S_r(x,\cl)\cap M_{x,r}\neq\emptyset\}.
\end{equation}
Thus, category $r$ is assigned to object $x$ if and only if at least one member of $M_{x,r}$ assigns $x$ to $r$.
When $M_{x,r}=M_x$ for every $x\in X$ and $r\in P$, the junta depends only on the object, and we obtain the \emph{\textbf{object-dependent junta rule}}
\begin{align}\label{eq:UJ-rule}
  J^{(M_x)}(\cl)(x)
  =
  \{c_{i}(x) \mid  i\in M_x\}.
\end{align}
Likewise, $M_{x,r}=M_r$ gives a rule $J^{(M_r)}$ whose junta depends only on the category, and $M_{x,r}=M$ gives the uniform junta rule $J^{(M)}=J^M$.

A social classification must assign at least one category to each object and leave no category empty.
The following proposition characterises when a junta rule is a CAC.

\begin{proposition}\label{thm:junta-common-member}
  Let $N$ be finite or infinite, $m\ge p\ge2$, and let $(M_{x,r})_{x\in X,r\in P}$ be a family of coalitions.
  Then $J^{(M_{x,r})}$ is a CAC if and only if the following conditions hold:
  \begin{subequations}\label{eq:pair-feasibility}
    \begin{align}
      \bigcap_{r\in P}M_{x,r}&\neq\emptyset\quad\text{for every }x\in X,\label{eq:pair-nonempty} \\
      \bigcap_{x\in X}M_{x,r}&\neq\emptyset\quad\text{for every }r\in P.\label{eq:pair-coverage}
    \end{align}
  \end{subequations}
  In particular, $J^{(M_x)}$ is a CAC if and only if
  \begin{align}\label{eq:common-member}
    \bigcap_{x\in X}M_x\neq\emptyset,
  \end{align}
  $J^{(M_r)}$ is a CAC if and only if $\bigcap_{r\in P}M_r\neq\emptyset$, and $J^M$ is a CAC if and only if $M\neq\emptyset$.
\end{proposition}

The two conditions above correspond to the two requirements on a social classification.
Condition \cref{eq:pair-nonempty} ensures that, at each object, some individual belongs to the junta for every category.
Whatever category she assigns to that object is therefore included in the social classification.
Without such an individual, everyone can assign the object to a category whose junta does not contain her, leaving the object without any category.

Condition \cref{eq:pair-coverage} ensures that, for each category, some individual belongs to its junta at every object.
Since her individual classification uses every category, she assigns at least one object to that category, and this assignment is included in the social classification.
Without such an individual, everyone can use the category only at an object whose corresponding junta does not contain her, leaving the category empty.
The individuals ensuring these requirements may differ across objects and categories.

By the definition of $\G_M$, the rule can be written in terms of its coalition families:
\begin{equation}\label{eq:J-via-G}
  r\in J^{(M_{x,r})}(\cl)(x)\iff S_r(x,\cl)\in\G_{M_{x,r}}.
\end{equation}
For $J^{(M_x)}$, fix an object $x$.
The support sets $\bigl(S_r(x,\cl)\bigr)_{r\in P}$ have union $N\in\G_{M_x}$.
By the grill property applied to this finite union, at least one support set belongs to $\G_{M_x}$, so the outcome at $x$ is nonempty whenever $M_x\neq\emptyset$.
Ensuring that no category is left empty additionally requires the common-member condition \cref{eq:common-member}.
Every individual in $\bigcap_{x\in X}M_x$ has her individual classification included in the social classification, and the assignments of the other individuals are included at the objects to whose juntas they belong.

The rule $F^{2'}$ of \cref{ex:running-junta} is $J^{(M_{x,r})}$ with $M_{x_2,b}=\{2\}$, $M_{x_4,c}=\{1\}$ and $M_{x,r}=N$ for every other pair.
Its juntas satisfy \cref{eq:pair-nonempty,eq:pair-coverage}, since voter~2 belongs to every junta at $x_2$ and to every junta for Biodiversity, and voter~1 belongs to every junta at $x_4$ and to every junta for Climate.
No voter, however, belongs to all juntas, and neither individual classification is included in $F^{2'}(\cl)$: voter~1's Biodiversity at $x_2$ and voter~2's Climate at $x_4$ are dropped.

Among the general junta rules that are CACs, only dictatorial rules are resolute.
For each $x$, choose $i_x\in\bigcap_{r\in P}M_{x,r}$, which is nonempty by \cref{eq:pair-nonempty}.
If some $M_{x,r}$ contained $j\neq i_x$, \cref{lem:real} would give a profile with $c_j(x)=r$ and $c_{i_x}(x)=s\neq r$, so both $r$ and $s$ would be accepted at $x$.
Thus $M_{x,r}=\{i_x\}$ for every $x,r$, and \cref{eq:pair-coverage} implies that all $i_x$ coincide.
This recovers the dictatorship of \citet{maniquet2016theorem} within the class of junta rules, and allowing multiple memberships is exactly what permits juntas with more than one member.

The next section characterises these rules and then returns to \cref{ex:running-quota} to ask whether individual responsibility and equal treatment allow any supported assignments to be rejected.

\section{Characterisations of junta rules}\label{sec:characterisation}

This section characterises junta rules by means of three axioms, each added to Independence and Supportiveness.
The three axioms -- Responsibility, Split Stability, and Rejection Stability -- express different requirements.
Responsibility looks for an individual whose report secures an assignment, Split Stability compares profiles before and after the support of a category is split, and Rejection Stability concerns coalitions that fail to secure one.
Split Stability and Rejection Stability imply the grill property, the former by splitting a winning coalition and the latter by combining losing ones, while Responsibility identifies the generating junta directly.

\subsection{Characterisation via Responsibility}\label{subsec:FRR}

We begin with a characterisation of junta rules based on the role of
individual reports in ensuring acceptance. Responsibility
requires that every accepted category with supporters be secured by
one supporter's complete classification, regardless of the other
individuals' reports.

Consider again the voters classifying proposals into
categories for the platform. Each voter assigns every proposal to
one category and uses every category, while the platform's classification may
assign several categories to the same proposal. These categories provide
different thematic groups for comparing proposals. One may require that every
accepted assignment with individual support be retained on the basis
of one supporter's complete classification, whatever the other
individuals report. This excludes acceptance that depends on agreement
among reports that are separately insufficient. It is an additional
normative requirement and does not follow from allowing multiple
memberships. \citet{fioravanti2025expert} motivates granting individuals
decisive power by their expertise or their direct connection to the
objects or categories involved. Responsibility makes no assumption about
individuals' expertise and does not require decision power to be
equally distributed among them.

For $i\in N$, denote by $(c_i,\bm d_{-i})$ the profile in which
individual $i$ reports $c_i$ and the other individuals report
$\bm d_{-i}\in\C^{N\setminus\{i\}}$.

\begin{axiom}[Responsibility]
  \label{ax:FRR}
  A CAC $F$ satisfies \emph{Responsibility} (Resp.) if, for every $\cl\in\C^N$, $x\in X$ and $r\in P$ such that $r\in F(\cl)(x)$ and $S_r(x,\cl)\neq\emptyset$, there exists $i\in S_r(x,\cl)$ such that
  $$
  r\in F(c_i,\bm d_{-i})(x)
  \qquad\text{for every }\bm d_{-i}\in\C^{N\setminus\{i\}}.
  $$
\end{axiom}

The individual may depend on the profile, the object and the category, but her report must ensure acceptance for every choice of the other reports.
Her report is sufficient for inclusion; it need not be necessary, and more than one individual may satisfy the requirement.
Under Supportiveness, every assignment can be traced to at least one such supporter, who in the participatory-budgeting setting could be asked to justify it.
This provides a basis for accountability, although Responsibility does not single out a unique supporter.

\begin{example}\label{ex:FRR-majority}
  Let $N=\{1,2,3\}$ be three trained classifiers assigning two proposal descriptions, $X=\{x_1,x_2\}$, to two thematic labels, $P=\{a,b\}$, read as Arts and Biodiversity.
  The only surjective classifications are $u=(a,b)$ and $v=(b,a)$.
  Let $F^5$ be majority voting at each object.
  This rule always uses both categories and therefore defines a CAC on this domain.
  At $\cl=(u,u,v)$, it assigns $\{a\}$ to $x_1$.
  The profiles $(u,v,v)$ and $(v,u,v)$ retain the reports of individuals~1 and~2, respectively, but both assign $\{b\}$ to $x_1$.
  Neither supporter's report ensures acceptance, so majority voting violates Responsibility, although each supporter is pivotal at $\cl$.
  $J^N$ accepts $a$ whenever either supporter retains $u$, whatever the others report.
\end{example}

Responsibility does not imply Supportiveness, since it imposes no condition on accepted categories without supporters; the constant CAC $F(\cl)(x)=P$ satisfies Responsibility.
When Independence also holds, the individual's assignment at the object suffices, whatever her assignments to the other objects.

\begin{lemma}\label{lem:local-responsibility}
  Let $F$ satisfy Independence and Responsibility.
  If $r\in F(\cl)(x)$ and $S_r(x,\cl)\neq\emptyset$, there exists $i\in S_r(x,\cl)$ such that
  \begin{equation}\label{eq:R}
    c'_i(x)=r\quad\Longrightarrow\quad r\in F(\cl')(x)
    \qquad\text{for every }\cl'\in\C^N.
  \end{equation}
\end{lemma}

\begin{proof}
  Take $i$ as in Responsibility and any $\cl'$ with $c'_i(x)=r$.
  The profile $(c_i,\cl'_{-i})$ belongs to $\C^N$, and Responsibility gives $r\in F(c_i,\cl'_{-i})(x)$.
  Since $c_i(x)=c'_i(x)$, this profile and $\cl'$ agree at $x$, and Independence gives $r\in F(\cl')(x)$.
\end{proof}

Without Independence the assignment alone need not suffice.
Under $F^3$ in \cref{ex:running-split}, Arts at $x_2$ is secured by voter~2's individual classification $c_2$, which uses Arts twice, but not by her assignment at $x_2$.
If she instead reports $c_2'=(c,a,b,c)$, she still assigns $x_2$ to Arts, yet the proposal leaves Arts.

In the single-valued setting, \citet[Remark~1]{fioravanti2025expert} calls an individual \emph{minimally semi-decisive} for a pair $(x,r)$ if her assignment of $x$ to $r$ suffices for society to assign $x$ to $r$ at every profile.
Condition \cref{eq:R} extends this notion to our setting.
The power concerns the inclusion of a category, which need not be the only category assigned to the object.
One voter may ensure that a proposal is placed in Biodiversity while another ensures that it is placed in Arts, and when they make these assignments both categories appear in the platform's classification.
Multiple memberships allow these powers to coexist at the same object.
For each pair $(x,r)$, let $M_{x,r}$ be the coalition of all individuals satisfying \cref{eq:R}.
Under Supportiveness, \cref{lem:local-responsibility} guarantees that every assigned category has a supporter in the corresponding coalition, and conversely the support of any member suffices.
For $F^{2'}$ in \cref{ex:running-junta}, these coalitions are the ones that define the rule: only voter~2 satisfies \cref{eq:R} for $x_2$ and Biodiversity, only voter~1 for $x_4$ and Climate, and both voters for every other pair.
This gives the following representation.

\begin{theorem}\label{thm:pair-representation}
  Let $m\geq p\geq2$.
  A CAC $F$ satisfies Independence, Supportiveness and Responsibility if and only if there exists a family $(M_{x,r})_{x\in X,r\in P}$ satisfying \cref{eq:pair-nonempty,eq:pair-coverage} such that
  \begin{equation}\label{eq:pair-representation}
    F(\cl)(x)=\{r\in P\mid S_r(x,\cl)\cap M_{x,r}\neq\emptyset\}\quad\text{for all }\cl\in\C^N\text{ and }x\in X.
  \end{equation}
  This family is unique.
\end{theorem}

Under Category Neutrality the same junta applies to all categories at a given object, and under Object Neutrality as well it applies to all objects.

\begin{corollary}\label{cor:FRR-junta}
  Let $m\geq p\geq2$.
  \begin{enumerate}[label=(\roman*),noitemsep]
    \item A CAC $F$ satisfies Independence, Supportiveness, Responsibility and Category Neutrality if and only if $F=J^{(M_x)}$ for a family of coalitions satisfying \cref{eq:common-member}.
    \item A CAC $F$ satisfies Independence, Supportiveness, Responsibility, Category Neutrality and Object Neutrality if and only if $F=J^M$ for a nonempty coalition $M\subseteq N$.
  \end{enumerate}
\end{corollary}

The proof of \cref{thm:pair-representation} identifies a responsible individual directly, so the characterisation also holds when $m=p$ or $p=2$.

\subsection{Characterisation via Split Stability}\label{subsec:SI}

Our second characterisation concerns changes in the support sets.
It asks how the social classification should respond when support for one category is divided between that category and a previously unused one.

Given $\cl\in\C^N$, $x\in X$, distinct categories $k,k'\in P$ and a nonempty coalition $B\subseteq S_k(x,\cl)$, let $\cl^{x,B\to k'}=(c'_i)_{i\in N}$ denote the profile obtained from $\cl$ by setting $c'_i(x)=k'$ for every $i\in B$ and leaving all other entries unchanged.
We call the passage from $\cl$ to $\cl^{x,B\to k'}$ a \emph{defection} of $B$ to $k'$ at $x$.
The defection is \emph{admissible} if $\cl^{x,B\to k'}\in\C^N$, that is, if $c_i^{-1}(k)\setminus\{x\}\neq\emptyset$ for every $i\in B$.
Not every defection from an arbitrary profile is admissible.
When $m>p$, however, \cref{lem:rich} implies that every opinion vector at $x$ is realised by a profile at which every defection at $x$ is admissible.

\begin{axiom}[Split Stability]
  \label{ax:SI}
  A CAC $F$ satisfies \emph{Split Stability} if, for all $x\in X$, all $\cl\in\C^{N}$, all $k,k'\in P$ such that $S_{k}(x,\cl)\neq\emptyset$ and $S_{k'}(x,\cl)=\emptyset$, and every nonempty $B\subseteq S_{k}(x,\cl)$ such that $\cl^{x,B\to k'} \in\C^N$, the following conditions hold:
  \begin{enumerate}[label=(SS.\arabic*),noitemsep]
    \item\label{cond:SI-i} $F\bigl(\cl^{x,B\to k'}\bigr)(x)\setminus\{k,k'\}=F(\cl)(x)\setminus\{k,k'\}$.
    \item\label{cond:SI-ii} $F\bigl(\cl^{x,B\to k'}\bigr)(x)\cap\{k,k'\}\neq\emptyset\iff k\in F(\cl)(x)$.
  \end{enumerate}
\end{axiom}

To understand the normative aspect of the condition, consider a profile at which some individuals assign $x$ to a category $k$ and \emph{nobody} assigns it to $k'$.
Suppose that part of the supporters of $k$ switch to $k'$, for reasons unrelated to manipulation: the two categories may be similar, or new information about $x$ may lead some individuals to revise their view.
Note that $x$ may also be classified into other categories under either profile.
Then, condition \ref{cond:SI-i} requires that the categories supported by others be unaffected by this change.
Condition \ref{cond:SI-ii} requires that at least one of $k$ and $k'$ be accepted after the split exactly when $k$ was accepted before it.
If the original support was strong enough to place $x$ in $k$, then after the split $x$ should be placed in $k$ or $k'$.
Conversely, if $x$ was not placed in $k$, the defectors should not be able to obtain a membership in $k$ or $k'$ by splitting.

We return to \cref{ex:running-split}.
\begin{example}[continues=ex:running-split]
  Suppose now that voter~2 moves $x_1$ from Arts to Climate, a category that nobody had proposed for $x_1$.
  Since she still uses Arts for $x_2$, her new report $c_2'=(c,a,b,c)$ is a classification, and the profile $\cl'=(c_1,c_2')$ gives
  \begin{align}
    F^3(\cl')=(\{a,c\},\{b\},\{c\},\{a,c\}).
  \end{align}
  At $x_1$, the support for Arts has been split between Arts and Climate.
  The proposal stays in Arts, enters Climate and remains outside Biodiversity.
  Under $F^1$, by contrast, the same change turns $\{a\}$ into $P$ at $x_1$, so Biodiversity appears although nobody proposed it.
  At $x_2$ and $x_4$, nobody changed her classification, yet the outcome changes from $\{a,b\}$ to $\{b\}$ and from $\{a\}$ to $\{a,c\}$, because voter~2 now uses Arts once and Climate twice. \hfill $\diamond$
\end{example}
When voter~2 moves $x_1$ from Arts to the previously unused category Climate, $F^3$ keeps $x_1$ under Arts, adds Climate and leaves Biodiversity out, as \ref{cond:SI-i} and \ref{cond:SI-ii} require.
The same move makes $F^1$ add Biodiversity at $x_1$, contrary to \ref{cond:SI-i}.
The changes that $F^3$ makes at $x_2$ and $x_4$ violate Independence. Split Stability constrains the outcome only at the object where the split occurs.

When $m=p$, every individual classification is a bijection, no admissible split exists, and the axiom is vacuous.
When $p=2$, no category is vacant unless the profile is unanimous at $x$; under Supportiveness, Split Stability then holds automatically.
We therefore state the result under $m>p\geq3$.

\begin{theorem}\label{thm:general_junta}
  Let $m>p\geq3$.
  A CAC $F$ satisfies Independence, Supportiveness and Split Stability if and only if there exists a family $(M_x)_{x\in X}$ of nonempty coalitions satisfying \cref{eq:common-member} such that $F=J^{(M_x)}$.
  The family is unique.
\end{theorem}

The proof, in \cref{app:proofs}, first shows that whether $k$ is assigned to $x$ depends only on the support set of $k$ at $x$, then that the family of winning support sets $\W_x$ does not depend on $k$, and finally that $\W_x$ is a grill.
Finiteness of $N$ is used only at the last step, to conclude that the grill is principal (\cref{prop:grill-structure}).
In particular, the junta at each object must be the same for all categories.
The rule $F^{2'}$ of \cref{ex:running-junta} shows what goes wrong otherwise: if voter~1 alone proposes Biodiversity for $x_2$, it is dropped, but if she moves $x_2$ to the previously unused category Climate instead, which is admissible when she uses Biodiversity for another proposal, the proposal enters Climate, contrary to \ref{cond:SI-ii}.
Adding Object Neutrality removes the dependence of the junta on the object.

\begin{corollary}\label{cor:uniform_junta}
  Let $m>p\geq3$.
  A CAC satisfies Independence, Supportiveness, Split Stability and Object Neutrality if and only if it is a uniform junta rule $J^M$ for some nonempty coalition $M$.
\end{corollary}

\subsection{Characterisation via Rejection Stability}\label{subsec:FRRC}

Our third characterisation starts from the idea that coalitions whose submitted classifications are not enough to secure an assignment should not become enough merely by being combined.
Consider again the platform's classification of proposals, and suppose that a proposal is not placed in Biodiversity at either of two profiles.
In each profile, the reports of the individuals who assign the proposal to Biodiversity are insufficient for the assignment.
Rejection Stability requires the assignment to remain rejected whenever each supporter in a third profile retains her own complete classification from one of the two rejecting profiles.
The requirement concerns complete classifications because, without Independence, an individual's assignments to other objects may affect the classification of the proposal.

\begin{axiom}[Rejection Stability]
  \label{ax:FRRC}
  A CAC $F$ satisfies \emph{Rejection Stability} (Rej. Stab.) if, for every $x\in X$, $r\in P$ and $\cl,\bm d,\bm e\in\C^N$ such that $r\notin F(\cl)(x)$ and $r\notin F(\bm d)(x)$, we have $r\notin F(\bm e)(x)$ whenever $e_i\in\{c_i,d_i\}$ for every $i\in S_r(x,\bm e)$.
\end{axiom}

The two rejecting profiles may coincide.
Each supporting report in $\bm e$ must be the same individual's report in $\cl$ or in $\bm d$, but different supporters may take their reports from different profiles, and only some of the original supporters need retain their support.
Individuals who do not support the assignment in $\bm e$ may submit arbitrary reports.

\begin{example}\label{ex:corroboration}
  Take the majority rule of \cref{ex:FRR-majority}, and consider $\cl=(u,v,v)$, $\bm d=(v,u,v)$ and $\bm e=(u,u,v)$.
  The outcome at $x_1$ is $\{b\}$ under both source profiles and $\{a\}$ under the target.
  Both supporters of $a$ in $\bm e$ retain their reports, since $e_1=c_1$ and $e_2=d_2$.
  Majority voting therefore violates Rejection Stability.
\end{example}

Rejection Stability does not imply Supportiveness either: the constant CAC $F(\cl)(x)=P$ satisfies it vacuously.
The closure in Rejection Stability extends to any finite number of rejecting profiles.

\begin{lemma}\label{lem:finite-families}
  Let $F$ satisfy Rejection Stability, and let $\cl^1,\ldots,\cl^K\in\C^N$, with $K\geq1$, be profiles such that $r\notin F(\cl^k)(x)$ for every $k$.
  If $\bm e\in\C^N$ satisfies $e_i\in\{c^1_i,\ldots,c^K_i\}$ for every $i\in S_r(x,\bm e)$, then $r\notin F(\bm e)(x)$.
\end{lemma}

\begin{proof}
  We argue by induction on $K$.
  For $K\leq2$, the claim is Rejection Stability.
  Let $K\geq3$, and let $U$ be the set of supporters $i\in S_r(x,\bm e)$ with $e_i\notin\{c^1_i,\ldots,c^{K-1}_i\}$, so that $e_i=c^K_i$ for $i\in U$.
  By \cref{lem:real}, there is $c^0\in\C$ with $c^0(x)\neq r$.
  Let $\bm e'$ be obtained from $\bm e$ by replacing $e_i$ with $c^0$ for every $i\in U$.
  Every supporter of $r$ at $x$ in $\bm e'$ retains a report from one of $\cl^1,\ldots,\cl^{K-1}$, so $r\notin F(\bm e')(x)$ by the induction hypothesis.
  Every supporter in $\bm e$ retains her report from $\bm e'$ or from $\cl^K$, and Rejection Stability gives $r\notin F(\bm e)(x)$.
\end{proof}

\begin{lemma}\label{lem:FRR-FRRC}
  Let $m\geq p\geq2$.
  \begin{enumerate}[label=(\roman*),noitemsep]
    \item On every population, Supportiveness and Responsibility imply Rejection Stability.
    \item If $N$ is finite, Rejection Stability implies Responsibility.
  \end{enumerate}
\end{lemma}

\begin{proof}
  (i) Let $F$ satisfy Supportiveness and Responsibility.
  Consider profiles $\cl,\bm d$ that reject $r$ at $x$ and a
  profile $\bm e$ satisfying the report condition of Rejection Stability.
  Suppose that $r\in F(\bm e)(x)$.
  By Supportiveness, $r$ has a supporter in $\bm e$, and Responsibility
  provides such a supporter $i$ whose report $e_i$ ensures
  acceptance for every choice of the other reports.
  Since $e_i=c_i$ or $e_i=d_i$, one of the source profiles must
  accept $r$ at $x$, a contradiction. Hence Rejection Stability holds.

  (ii) Let $N$ be finite and suppose that $F$ satisfies Rejection Stability.
  Take $\bm e$, $x$ and $r$ such that $r\in F(\bm e)(x)$ and
  $S_r(x,\bm e)\neq\emptyset$.
  Suppose that Responsibility fails at this profile, object and category.
  Then, for every $i\in S_r(x,\bm e)$, there is a profile
  $\bm d^i\in\C^N$ such that
  $$
  d^i_i=e_i
  \qquad\text{and}\qquad
  r\notin F(\bm d^i)(x).
  $$
  These finitely many profiles reject $r$ at $x$, and every
  supporter $i$ in $\bm e$ retains her report from $\bm d^i$.
  By \cref{lem:finite-families}, $r\notin F(\bm e)(x)$,
  a contradiction. Thus Responsibility holds.
\end{proof}

Thus, on finite populations, Responsibility and Rejection Stability are equivalent under Supportiveness, without Independence or either neutrality axiom.
Combining \cref{lem:FRR-FRRC} with \cref{thm:pair-representation,cor:FRR-junta} gives the third characterisation.

\begin{theorem}\label{thm:frrc-characterisation}
  Let $m\geq p\geq2$.
  \begin{enumerate}[label=(\roman*),noitemsep]
    \item A CAC $F$ satisfies Independence, Supportiveness and Rejection Stability if and only if $F=J^{(M_{x,r})}$ for a unique family of coalitions satisfying \cref{eq:pair-nonempty,eq:pair-coverage}.
    \item Adding Category Neutrality characterises the object-dependent junta rules $J^{(M_x)}$ satisfying \cref{eq:common-member}, and adding Object Neutrality as well characterises the uniform junta rules $J^M$ with $M\neq\emptyset$.
  \end{enumerate}
\end{theorem}

Under Independence and Supportiveness and when $m>p\geq3$, Split Stability and Rejection Stability both yield the grill property.
Split Stability requires that splitting a winning coalition leaves at least one winning part, while Rejection Stability requires that combining two losing coalitions never produces a winning one.
The difference is that Split Stability also forces the winning coalitions to be the same for all categories at a given object, whereas Rejection Stability allows them to depend on the category.
Thus, on finite populations, \cref{thm:general_junta,thm:frrc-characterisation} show that Split Stability is equivalent to Rejection Stability together with Category Neutrality.

Independence cannot be omitted.
The rule $F^3$ of \cref{ex:running-split} satisfies Supportiveness and both neutrality axioms.
Every category it accepts is secured by the complete report of voter~1 or of voter~2, so Responsibility holds, and \cref{lem:FRR-FRRC} gives Rejection Stability.
Yet it is not a junta rule, since its outcomes at $x_2$ and $x_4$ change although nobody changes her classification of these proposals.
Part (ii) of \cref{lem:FRR-FRRC} fails on infinite populations, as \cref{sec:infinite} shows.

\subsection{Relations between the axioms}\label{subsec:role-supp}\label{subsubsec:FRR}\label{subsubsec:grills}

On finite populations, under Independence and Supportiveness and for $m>p\geq3$, Split Stability is equivalent to Responsibility together with Category Neutrality, and to Rejection Stability together with Category Neutrality (\cref{thm:general_junta,cor:FRR-junta,lem:FRR-FRRC}).
Rejection Stability implies Responsibility for all $m\geq p\geq2$, even without Independence or Supportiveness (\cref{lem:FRR-FRRC}(ii)).
We now examine whether Unanimity can replace Supportiveness in the characterisations and what happens when we impose Anonymity.

\subsubsection{Replacing Supportiveness with Unanimity}\label{subsubsec:supportiveness}

Our characterisations impose Supportiveness in place of the Unanimity axiom of \citet{maniquet2016theorem}.
The following proposition identifies when Unanimity and the other axioms already imply Supportiveness.

\begin{proposition}\label{prop:una-replace}
  Let $m\geq p\geq 2$.
  \begin{enumerate}[label=(\roman*),noitemsep]
    \item If $m>p$, every CAC satisfying Unanimity, Independence and Split Stability satisfies Supportiveness.
    \item Every CAC satisfying Unanimity and Rejection Stability satisfies Supportiveness.
  \end{enumerate}
\end{proposition}

Thus \cref{thm:general_junta,cor:uniform_junta,thm:frrc-characterisation} remain valid when Supportiveness is replaced by Unanimity.
For the characterisations through Responsibility in \cref{thm:pair-representation,cor:FRR-junta}, this replacement is valid if and only if $p=2$ (cf.\ \cref{prop:oligarchy-resolution}).

For $p\geq3$, Supportiveness is therefore needed as a separate assumption in the characterisation through Responsibility.
Responsibility, as stated, imposes no condition on accepted categories without supporters.
In \cref{ex:running-oligarchy}, Climate at $x_2$ is such a category: Responsibility imposes nothing on it, and Unanimity does not exclude it because the voters disagree.
One could strengthen Responsibility by requiring every accepted category to be secured by some individual's complete classification, whether or not that individual supports the assignment.
Combined with Unanimity, this condition forces the responsible individual to be a supporter: if she assigned $x$ to some $s\neq r$, letting every other individual do the same would produce a profile that is unanimous on $s$ at $x$.
Unanimity together with this condition is therefore equivalent to Supportiveness together with Responsibility, and we keep the more transparent second combination.

\subsubsection{Anonymity and Responsibility}\label{subsubsec:anon-FRR}

The preceding characterisations allow individuals to differ in their authority to secure an assignment.
We now consider the effect of requiring equal treatment.

\begin{axiom}[Anonymity]
  \label{ax:anon}
  A CAC $F$ satisfies \emph{Anonymity} if, for every bijection $\tau\colon N\to N$ and every $\cl\in\C^N$, $F((c_{\tau(i)})_{i\in N})=F(\cl)$.
\end{axiom}

Note that in the running example, only $F^{2'}$ and $F^3$ violate Anonymity.
In \cref{ex:running-junta}, if the voters exchange their reports, the outcome at $x_2$ becomes $\{a,b\}$ instead of $\{a\}$; and
in \cref{ex:running-split}, only voter~2's categories are subject to the repetition requirement.

Among general junta rules that are CACs, Anonymity holds if and only if $M_{x,r}=N$ for every object-category pair.
Indeed, if one individual's sole support secures an assignment, Anonymity gives every individual the same authority.
The preceding characterisations therefore yield the following result.

\begin{corollary}\label{cor:anon-finite}
  Let $N$ be finite and $m\geq p\geq2$.
  A CAC satisfies Independence, Supportiveness, Anonymity and either Responsibility or Rejection Stability (or Split Stability, when $m>p\geq3$) if and only if it is $J^N$.
\end{corollary}

Under the maintained axioms, Anonymity and Responsibility are thus compatible only through the whole-population junta rule $J^N$, which accepts every supported assignment.
To retain the possibility of rejecting some supported assignments while preserving Responsibility, individuals must be allowed to differ in whose support suffices.
The quota procedure in \cref{ex:running-quota} treats individuals equally but violates Responsibility, since neither voter can secure an assignment for which both voters' support is required.
Allowing different individuals to have different decision power therefore permits junta rules to reject some supported assignments.\footnote{\citet{CaillouxEtAl2025} showed that, in the resolute setting, no anonymous and neutral rules exist for many sizes of the problem even when Independence is not imposed.
  Meanwhile, the whole-population junta rule $J^N$ satisfies Anonymity, Independence, Supportiveness, Object Neutrality, and Category Neutrality for all numbers of individuals, objects and categories.
}

\Cref{tab:examples} collects the properties of the running examples.
In particular, $F^{2'}$ and $F^4$ illustrate the distinction above: the former satisfies Responsibility and the latter satisfies Anonymity, although they give the same outcome at the original profile.
The table also includes the majority rule $F^5$ of \cref{ex:FRR-majority}.

\begin{table}[t]
  \centering
  \caption{Axioms satisfied ($\checkmark$) and violated ($\times$) by the rules in \cref{ex:running-oligarchy,ex:running-junta,ex:running-split,ex:running-quota,ex:FRR-majority}.}
  \label{tab:examples}
  \small
  \begin{tabular}{lcccccc}
    \toprule
    & $F^1$ & $F^2$ & $F^{2'}$ & $F^3$ & $F^4$ & $F^5$ \\
    \midrule
    Independence          & $\checkmark$ & $\checkmark$ & $\checkmark$ & $\times$     & $\checkmark$ & $\checkmark$ \\
    Unanimity             & $\checkmark$ & $\checkmark$ & $\checkmark$ & $\checkmark$ & $\checkmark$ & $\checkmark$ \\
    Supportiveness        & $\times$     & $\checkmark$ & $\checkmark$ & $\checkmark$ & $\checkmark$ & $\checkmark$ \\
    \midrule
    Split Stability       & $\times$     & $\checkmark$ & $\times$     & $\checkmark$ & $\times$     & $\checkmark^{\dagger}$ \\
    Responsibility                   & $\checkmark$ & $\checkmark$ & $\checkmark$ & $\checkmark$ & $\times$     & $\times$ \\
    Rejection Stability                  & $\times$     & $\checkmark$ & $\checkmark$ & $\checkmark$ & $\times$     & $\times$ \\
    \midrule
    Object Neutrality     & $\checkmark$ & $\checkmark$ & $\times$     & $\checkmark$ & $\times$     & $\checkmark$ \\
    Category Neutrality   & $\checkmark$ & $\checkmark$ & $\times$     & $\checkmark$ & $\times$     & $\checkmark$ \\
    Anonymity             & $\checkmark$ & $\checkmark$ & $\times$     & $\times$     & $\checkmark$ & $\checkmark$ \\
    \bottomrule
  \end{tabular}
  \par\smallskip
  \begin{minipage}{0.7\textwidth}
    {\footnotesize \emph{Notes:} $F^1$ and $F^2$ correspond to $O^N_+$ and $J^N$, respectively; $F^{2'}$ is a general junta rule; $F^4$ is a quota rule; and $F^5$ is the majority rule on the domain $n=3$, $m=p=2$ used in \cref{ex:FRR-majority}.\\ $^{\dagger}$Vacuously, since no split is admissible when $m=p$.}
  \end{minipage}
\end{table}

\section{Infinite populations}\label{sec:infinite}

On finite populations, when $m>p\geq3$, the three axioms of \cref{sec:characterisation} are interchangeable under Independence, Supportiveness and Category Neutrality.
This section shows that they come apart when the population is infinite.
The question is natural because the theorem of \citet{maniquet2016theorem}, our benchmark, fails on an infinite population (\cref{sec:inf-mm}).
Split Stability and Rejection Stability make the winning coalitions a grill that need not be principal and characterise a strictly larger class, the limiting junta rules (\cref{sec:inf-grill}).
Responsibility keeps the grill principal and characterises the junta rules as before (\cref{sec:inf-frr}).
Throughout, $N$ is an arbitrary set with at least two members, finite or infinite, while $X$ and $P$ remain finite with $m \ge p \ge 2$.

\subsection{Benchmark: the Maniquet--Mongin theorem with an infinite population}\label{sec:inf-mm}

\citet{maniquet2016theorem} show that the decisive coalitions of a resolute CAC satisfying Independence and Unanimity form an ultrafilter $\U$ on $N$, with $F(\cl)(x)=\{r\}$ if and only if $S_{r}(x,\cl) \in \U$.
Their argument uses the finiteness of $N$ only through the fact that every ultrafilter on a finite set is principal.

\begin{corollary}\label{cor:mm-possibility}
  Assume the axiom of choice.
  If $N$ is infinite, there exists a resolute CAC satisfying Independence, Supportiveness and Unanimity that is not dictatorial.
\end{corollary}

\begin{proof}
  Let $\U$ be a free ultrafilter on $N$ and let $F(\cl)(x)=\{r\}$ for the unique $r \in P$ with $S_{r}(x,\cl) \in \U$; uniqueness and existence follow from \ref{p:filter}, \ref{p:emptyset} and the grill property of $\U$ applied to the finite partition $(S_r(x,\cl))_{r\in P}$.
  Since $X$ is finite, $\bigcap_{x \in X} S_{r_x}(x,\cl) \in \U$, where $F(\cl)(x)=\{r_x\}$, and any individual $i$ in this intersection satisfies $c_i(x)=r_x$ for every $x$, so the outcome is surjective.
  Independence and Supportiveness are immediate, and a dictator $d$ would give $\{d\} \in \U$, contradicting freeness.
\end{proof}

Hence the theorem of Maniquet and Mongin holds if and only if $N$ is finite.
The rule in \cref{cor:mm-possibility} relies on a free ultrafilter, whose existence requires a weak form of the axiom of choice, so it cannot be defined explicitly.
This is the classification counterpart of the observation of \citet{fishburn1970arrow} and \citet{KirmanSondermann1972}.

\subsection{Split Stability and Rejection Stability: limiting junta rules}\label{sec:inf-grill}

Fix $x \in X$ and $r \in P$.
A coalition $A \subseteq N$ is \emph{winning for $r$ at $x$} if $r \in F(\cl)(x)$ for every $\cl \in \C^{N}$ with $S_{r}(x,\cl) = A$, and \emph{null for $r$ at $x$} if $r \notin F(\cl)(x)$ for every such $\cl$.
We write $\W_{x,r}$ for the family of winning coalitions.

Under Independence and Supportiveness, Split Stability and Rejection Stability each make every coalition either winning or null for $r$ at $x$, and make $\W_{x,r}$ a grill, whether $N$ is finite or infinite (\cref{lem:grill} in \cref{app:proofs}).
Under Split Stability, or under Rejection Stability and Category Neutrality, $\W_{x,r}=\W_x$ does not depend on $r$, and under Object Neutrality as well it does not depend on $x$.
On a finite population this grill is principal, which yields the junta rules of \cref{sec:characterisation}.
On an infinite population it need not be.
To describe the resulting rules, call a nonempty family $\mathcal{M}$ of nonempty coalitions \emph{directed} if for all $M, M' \in \mathcal{M}$ there is $M'' \in \mathcal{M}$ with $M'' \subseteq M \cap M'$.

\begin{definition}\label{def:limiting-junta}
  Given a family $(\mathcal{M}_{x})_{x \in X}$ of directed families, the \emph{\textbf{object-dependent limiting junta rule}} $L^{(\mathcal{M}_{x})}$ is defined, for all $\cl \in \C^{N}$ and all $x \in X$, by
  \begin{align}\label{eq:LJ-rule}
    L^{(\mathcal{M}_{x})}(\cl)(x)
    = \bigcap_{M \in \mathcal{M}_{x}} J^{M}(\cl)(x)
    = \bigl\{ r \in P \mid S_{r}(x,\cl) \cap M \neq \emptyset \text{ for all } M \in \mathcal{M}_{x} \bigr\}.
  \end{align}
  It is a \emph{uniform limiting junta rule}, written $L^{\mathcal{M}}$, if $\mathcal{M}_{x} = \mathcal{M}$ for every $x \in X$.
\end{definition}

Reading each $M \in \mathcal{M}_{x}$ as a committee, a category is attached to $x$ exactly when every committee contains someone who attaches it.
The rule is the limit of junta rules as the committees shrink along $\mathcal{M}_{x}$, in the same way as the limiting dictatorial rules of \citet{torres2005limiting} are limits of dictatorships.
An object-dependent junta rule $J^{(M_{x})}$ is the case $\mathcal{M}_{x} = \{M_{x}\}$.

For each $x$, the family $\G(\mathcal{M}_x)=\{S\subseteq N\mid S\cap M\neq\emptyset\text{ for all }M\in\mathcal M_x\}$ is a grill, every grill arises in this way from the directed family of its blocking coalitions, and $r\in L^{(\mathcal M_x)}(\cl)(x)$ if and only if $S_r(x,\cl)\in\G(\mathcal M_x)$ (\cref{lem:limiting-cac} in \cref{app:proofs}).
The limiting junta rules are therefore exactly the rules that accept a category when its support set belongs to a grill.
Such a rule is a CAC if and only if
\begin{equation}\label{eq:selection}
  \bigcap_{x\in X}M_x\neq\emptyset\quad\text{for every family }(M_x)_{x\in X}\text{ with }M_x\in\mathcal M_x\text{ for each }x,
\end{equation}
a condition that holds automatically for uniform limiting junta rules.
This yields the characterisation on an arbitrary population.

\begin{theorem}\label{thm:limiting}
  Let $N$ be finite or infinite.
  \begin{enumerate}[label=(\alph*),noitemsep]
    \item Let $m>p\geq3$. A CAC $F$ satisfies Independence, Supportiveness and Split Stability if and only if $F=L^{(\mathcal M_x)}$ for a family of directed families satisfying \cref{eq:selection}.
    \item Let $m\geq p\geq2$. A CAC $F$ satisfies Independence, Supportiveness, Rejection Stability and Category Neutrality if and only if $F=L^{(\mathcal M_x)}$ for a family of directed families satisfying \cref{eq:selection}.
  \end{enumerate}
  In both cases, adding Object Neutrality characterises the uniform limiting junta rules.
\end{theorem}

Thus, when $N$ is finite or infinite and $m>p\geq3$, a CAC satisfying Independence and Supportiveness satisfies Split Stability if and only if it satisfies Rejection Stability and Category Neutrality.
The next proposition compares limiting junta rules with junta rules.

\begin{proposition}\label{prop:lj-vs-junta}
  \begin{enumerate}[label=(\roman*),noitemsep]
    \item $L^{(\mathcal{M}_{x})}=J^{(M_x)}$ for some family $(M_x)_{x\in X}$ of coalitions if and only if every $\mathcal{M}_{x}$ has a smallest member, in which case $M_x$ is the smallest member of $\mathcal{M}_{x}$.
    \item If $N$ is infinite and $\mathcal{M}$ is the family of cofinite coalitions, then $L^{\mathcal{M}}(\cl)(x) = \{ r \in P \mid S_{r}(x,\cl) \text{ is infinite} \}$, and $L^{\mathcal{M}}$ is not a junta rule: $L^{\mathcal{M}}\neq J^{(M_{x,r})}$ for every family $(M_{x,r})_{x\in X,r\in P}$ of coalitions.
  \end{enumerate}
\end{proposition}

By \cref{lem:duality}, part (i) says that a limiting junta rule is an object-dependent junta rule exactly when every winning family $\W_x$ is principal.
On a finite population this is always the case: by directedness, a member of $\mathcal M_x$ of minimal cardinality is contained in every other member, so every limiting junta rule is an object-dependent junta rule.
This is the counterpart of the fact that every ultrafilter on a finite set is principal, and it is the only point at which finiteness matters.
Under the respective assumptions of \cref{thm:limiting}, Split Stability and Rejection Stability therefore characterise a class strictly larger than the object-dependent junta rules when $N$ is infinite.
We call the rule in (ii) the \emph{Fr\'echet rule}, since its winning family consists of the infinite sets, which are exactly the sets that meet every member of the Fr\'echet filter of cofinite sets.\footnote{The Fr\'echet filter on $N$ is $\{A\subseteq N\mid N\setminus A\text{ is finite}\}$.}
It satisfies Independence, Supportiveness, Split Stability, Rejection Stability and both neutrality axioms, yet it is not a junta rule.
Its committees are defined without any appeal to the axiom of choice, so its existence is provable in ZF, whereas that of the rules in \cref{cor:mm-possibility} is not.
The difference lies in the form of the output.
A set-valued rule needs only a directed family of committees, whereas a single-valued rule needs its decisive coalitions to form an ultrafilter.
The Fr\'echet rule also violates Responsibility: starting from unanimous support for $r$ at $x$, retain any one individual's report and let all others assign $x$ to another category, which is possible by \cref{lem:real}; the category then has a single supporter and is rejected.

\subsection{Responsibility: junta rules on every population}\label{sec:inf-frr}

Responsibility continues to characterise junta rules on infinite populations.

\begin{corollary}\label{cor:FRR-infinite}
  Let $N$ be infinite and $m\geq p\geq2$.
  Then \cref{thm:pair-representation} and \cref{cor:FRR-junta} remain valid.
  In particular, an object-dependent limiting junta rule satisfies Responsibility if and only if every winning family $\W_x$ is principal, that is, if and only if it is an object-dependent junta rule.
\end{corollary}

In fact, the proofs of \cref{thm:pair-representation} and \cref{cor:FRR-junta} do not use the finiteness of $N$.
An object-dependent limiting junta rule satisfies Independence, Supportiveness and Category Neutrality, so by \cref{cor:FRR-junta}(i) it satisfies Responsibility if and only if it is an object-dependent junta rule.
By \cref{prop:lj-vs-junta}(i) and \cref{lem:duality}, this holds if and only if every $\W_x$ is principal.

The contrast with \cref{thm:limiting} lies in the way each axiom identifies the members of a junta.
Split Stability and Rejection Stability relate a coalition to two of its parts at a time: if the union of two coalitions is winning, so is one of them (\cref{lem:grill}).
On a finite population, splitting a winning coalition again and again ends with a single winning individual, and these individuals form the junta.
On an infinite population the process need not end.
At every accepted category, Responsibility names a supporter whose report secures the category on her own (\cref{lem:local-responsibility}), so each member of the junta is identified directly, whatever the size of the population.
Put differently, under Supportiveness Responsibility requires rejection to survive the combination of any number of rejecting profiles, one for each supporter, whereas Rejection Stability requires this only for finitely many (\cref{lem:finite-families}), and the two coincide when $N$ is finite (\cref{lem:FRR-FRRC}).

\section{Concluding remarks}\label{sec:conclusion}

This paper asks how little of the framework of \citet{maniquet2016theorem} must be given up to escape their impossibility.
We keep their individual domain, in which every individual reports a single category for each object and uses every category, together with Independence and the requirement that the social classification classify every object and leave no category empty.
We allow an object to belong to several categories in the social classification and strengthen Unanimity to Supportiveness.
In the resolute benchmark, Independence and Unanimity already imply Supportiveness (\cref{prop:benchmark}).
With at least three categories, Independence and Unanimity permit a set-valued oligarchy rule that places every disputed object in all categories, including unsupported ones (\cref{prop:oligarchy-resolution}).

On finite populations, Independence and Supportiveness together with either Responsibility or Rejection Stability characterise the general junta rules $J^{(M_{x,r})}$ (\cref{thm:pair-representation,thm:frrc-characterisation}).
Together with Split Stability they characterise the object-dependent junta rules $J^{(M_x)}$ when $m>p\geq3$ (\cref{thm:general_junta}), and the neutrality axioms reduce the juntas to a single coalition (\cref{cor:FRR-junta,cor:uniform_junta,thm:frrc-characterisation}).
In each case the juntas are uniquely determined by the rule.
Supportiveness can be replaced by Unanimity in the characterisations through Split Stability and Rejection Stability but, when $p\geq3$, not in the one through Responsibility (\cref{prop:una-replace}).
When outcomes are required to be singletons, every junta shrinks to the same individual, and the dictatorship of Maniquet and Mongin reappears as the resolute special case.

The characterisations describe what multiple memberships make possible; they do not single out a recommended junta.
Adding Anonymity leaves only $J^N$, which accepts every proposed membership (\cref{cor:anon-finite}).
Under the maintained axioms, a rule that rejects some supported memberships while satisfying Responsibility must therefore give some individuals more authority than others, whereas a rule that is selective and treats individuals equally, such as the quota procedure of \cref{ex:running-quota}, must give up Responsibility.

\begin{table}[!t]
  \centering
  \caption{Coalitional structures and rules in preference aggregation and in classification aggregation.}
  \label{tab:dual}
  \footnotesize
  \setlength{\tabcolsep}{4pt}
  \renewcommand{\arraystretch}{1.05}

  \begin{tabularx}{\textwidth}{@{}>{\raggedright\arraybackslash}p{2.4cm}>{\raggedright\arraybackslash}p{2.25cm}>{\raggedright\arraybackslash}X>{\raggedright\arraybackslash}p{2.175cm}>{\raggedright\arraybackslash}X>{\raggedright\arraybackslash}p{2.175cm}@{}}
    \toprule
    & & \multicolumn{2}{c}{Finite $N$} & \multicolumn{2}{c}{Infinite $N$} \\
    \cmidrule(lr){3-4}\cmidrule(l){5-6}
    Social outcome & Axioms & Structure; rules & Source & Structure; rules & Source \\
    \midrule

    \multicolumn{6}{@{}l}{\emph{\textbf{Preference aggregation}}} \\
    \addlinespace

    Complete, transitive
    & WP, IIA
    & Ultrafilter {\scriptsize (= \emph{Filter} $\cap$ Grill)}; dictatorship
    & \citet{KirmanSondermann1972}
    & Ultrafilter, possibly free; non-dictatorial rules possible under AC
    & \citet{fishburn1970arrow}, \citet{KirmanSondermann1972} \\

    Complete, quasi-transitive
    & WP, IIA
    & Principal filter; oligarchy
    & \citet{Gibbard2014}
    & Filter, possibly nonprincipal; non-oligarchic rules possible
    & \citet{Hansson1976} \\

    Transitive, possibly incomplete
    & WP, IIA
    & Filter (decisive)/ Grill (quasi-decisive); oligarchy
    & \citet{Weymark1984}, \citet{cato2013quasi}
    & Filter (decisive)/ grill (quasi-decisive); not necessarily principal
    & \citet{BossertSuzumura2009}, \citet{cato2013quasi} \\

    \midrule

    \multicolumn{6}{@{}l}{\emph{\textbf{Classification aggregation}}} \\
    \addlinespace

    Single-valued \& surjective
    & I, Unanimity
    & Ultrafilter {\scriptsize (= Filter $\cap$ \emph{Grill})}; dictatorship
    & \citet{maniquet2016theorem}
    & Ultrafilter, possibly free; non-dictatorial rules possible under AC
    & Corollary~\ref{cor:mm-possibility} \\
    \addlinespace

    Set-valued \& covering
    & I, Supportiveness, SS
    & Principal grill; $J^{(M_x)}$
    & Theorem~\ref{thm:general_junta}
    & Grill; $L^{(\mathcal M_x)}$
    & Theorem~\ref{thm:limiting} \\

    \textit{ditto}
    & I, Supportiveness, Rej. Stab., CN
    & Principal grill; $J^{(M_x)}$
    & Theorem~\ref{thm:frrc-characterisation}
    & Grill; $L^{(\mathcal M_x)}$
    & Theorem~\ref{thm:limiting} \\

    \textit{ditto}
    & I, Supportiveness, Resp., CN
    & Principal grill; $J^{(M_x)}$
    & Corollary~\ref{cor:FRR-junta}
    & Principal grill; $J^{(M_x)}$
    & Corollary~\ref{cor:FRR-infinite} \\

    \bottomrule
  \end{tabularx}

  \par\smallskip
  \begin{minipage}{\textwidth}
    \scriptsize
    \emph{Notes.}
    In preference aggregation, individual preferences are complete and transitive, the domain is unrestricted, and the alternative set $X$ contains at least three elements.
    The filter and grill statements concern families of decisive and quasi-decisive coalitions, respectively, and do not by themselves uniquely determine the aggregation rule.
    The structures called quasi-ultrafilters by \citet{cato2013quasi} are termed grills here.
    WP denotes weak Pareto, IIA independence of irrelevant alternatives, and AC the axiom of choice.
    In classification aggregation, I denotes Independence, SS Split Stability, and CN Category Neutrality.
    The single-valued classification row is stated for $m\geq p\geq3$.
    The Split Stability characterisations require $m>p\geq3$, while the Responsibility and Rejection Stability characterisations hold for $m\geq p\geq2$.
    The junta and limiting junta rules satisfy the covering conditions stated in the corresponding results.
  \end{minipage}

\end{table}

\Cref{tab:dual} places these results alongside the decisive structures of preference aggregation.
An ultrafilter is exactly a family that is both a filter and a grill (\cref{fact:filter_grill}).
The Arrovian literature on weakened collective rationality retains the filter half: under quasi-transitive social preferences, the decisive coalitions form the principal filter $\F_M$ of an oligarchy $M$, and each oligarch can veto a strict social preference \citep{Gibbard2014}.
Multiple memberships retain the grill half: for each object and category, the coalitions that secure the membership form the principal grill $\G_{M_{x,r}}$, and each member of the junta can secure the inclusion of the category she proposes.
The two structures are dual (\cref{lem:duality}), since the coalitions that can block a strict social preference under the oligarchy with oligarchs $M$ are exactly those that secure a membership under $J^M$.
If the coalitions able to secure a membership formed a filter $\F_M$ with $|M|\geq2$, an object on which two members of $M$ disagree would receive no category (\cref{subsec:translating}).
Multiple memberships let the powers of different junta members coexist at the same object, and Supportiveness ensures that every membership they create rests on an individual report.

On infinite populations the three axioms come apart.
The theorem of Maniquet and Mongin fails there, as Arrow's theorem does \citep{fishburn1970arrow,KirmanSondermann1972}, but only through free ultrafilters, whose existence requires a weak form of the axiom of choice (\cref{cor:mm-possibility}).
Split Stability, and Rejection Stability together with Category Neutrality, characterise the limiting junta rules, whose winning coalitions form grills that need not be principal (\cref{thm:limiting}).
This class contains the Fr\'echet rule, which assigns a category to an object exactly when infinitely many individuals propose it and is defined without any appeal to choice (\cref{prop:lj-vs-junta}).
Responsibility, by contrast, keeps the grills principal on every population (\cref{cor:FRR-infinite}), because it identifies each member of a junta directly instead of reaching her by splitting coalitions.
The axioms thus differ in whether a membership must be secured by an individual supporter or may rest on infinite groups, none of whose finite subgroups secures it on its own.

Equal treatment also fares differently than in Arrovian aggregation.
On infinite populations, anonymity is inconsistent with Arrow's unanimity and independence conditions on sufficiently rich domains \citep{mihara1997anonymity}.
The same is true of resolute CACs satisfying Independence and Unanimity, whose decisive coalitions form an ultrafilter (\cref{sec:inf-mm}), since no ultrafilter on $N$ is invariant under all permutations of $N$.
With multiple memberships, by contrast, $J^N$ satisfies Anonymity on every population, and the Fr\'echet rule does so on every infinite population, because permutations of $N$ preserve whether a support set is nonempty or infinite.

We close with three directions for future research.

In our model individuals report one category for each object, so multiple memberships can arise only in the social classification.
In many applications, however, individuals may themselves wish to place an object in several categories; a voter may judge that a project belongs to Biodiversity and Climate.
A natural extension lets individual reports take the same form as social classifications, so that the input and output domains coincide.
Supportiveness then requires every collective membership to be proposed by some individual, and an independence condition applied separately to each membership statement, combined with Unanimity, already delivers it.
The junta rules remain well defined and are CACs under the same conditions on their juntas.
Moreover, since single-valued reports remain admissible, any rule satisfying extensions of our axioms that reduce to them on single-valued profiles must coincide with a junta rule on those profiles.
An individual may now support several categories at the same object, so the support sets at an object overlap instead of partitioning the population.
Responsibility and Rejection Stability are stated in terms of individual reports and extend almost verbatim, whereas Split Stability, which moves supporters from one category to another, admits several non-equivalent reformulations once an individual can add a category without abandoning another.
Whether these axioms still deliver grills on the richer domain, and how the rules they characterise behave at profiles with set-valued reports, remain open.

We have not studied incentives.
Under a junta rule, an individual's report at an object matters only through the category she proposes there, and only if she belongs to the junta for that object and category.
She can add that category to the social classification, but she cannot remove a category proposed by another member of the corresponding junta.
How this limited influence translates into incentives depends on how individuals evaluate social classifications with multiple memberships, for instance by their proximity to their own reports, so a strategic analysis must first specify such evaluations.
Under the Fr\'echet rule, changing the reports of finitely many individuals never changes whether a support set is infinite, so no finite coalition can affect the outcome, while infinite coalitions can; on infinite populations, the relevant notion of manipulation is therefore coalitional.

We have adopted the membership reading throughout.
Under the tie reading, the set assigned to an object lists the categories that remain admissible for it, and a single category is selected later.
The natural requirement on outcomes is then that some selection be surjective, which means that, for every set $T$ of categories, at least $|T|$ objects are assigned a set that meets $T$.
The oligarchy rule $O^M_{+}$ satisfies this requirement, since the classification of any member of $M$ is a surjective selection, and under this reading it becomes a reasonable counterpart of an oligarchy; junta rules may fail it, as $F^{2'}$ in \cref{ex:running-junta} does.
As noted in \cref{sec:notation}, if the final selection satisfies the requirements of \citet{maniquet2016theorem}, their impossibility returns at that stage.
The question is therefore what the set-valued stage can achieve by itself, and in particular whether the filter-based structures that are unavailable under the membership reading reappear under the tie reading, as they do in preference aggregation.
A related question concerns populations modelled by measure spaces \citep{Cato2021}, in which individuals may be negligible, and in which the contrast between Responsibility, which relies on individual supporters, and the two stability axioms, which only compare coalitions, can be expected to take a different form.

\newpage
\setlength{\bibsep}{0pt}
\bibliography{ref}
\newpage
\appendix

\section{Proofs}\label{app:proofs}

\subsection*{Proof of \texorpdfstring{\cref{prop:oligarchy-resolution}}{Proposition}}

\begin{enumerate}[label=(\roman*),leftmargin=*,noitemsep]
  \item Fix $i\in M$. By definition, $c_i(x)\in O^M_+(\cl)(x)$ at every object.
  Thus outcomes are nonempty, and every category is used because $c_i$ is surjective.
  The outcome at $x$ depends only on the reports at $x$, and unanimous support for $r$ gives $\{r\}$.
  Hence Independence and Unanimity hold.
  The definition is unchanged by relabelling objects or categories, so both neutrality axioms hold.

  \item If $M=\{d\}$, the rule returns $\{c_d(x)\}$ at every object.
  If $|M|\geq2$, \cref{lem:real} gives a profile where two members disagree at some object.
  The outcome there is $P$, so the rule is not dictatorial.

  \item If the members of $M$ agree on $r$, both rules return $\{r\}$.
  If they disagree and $p=2$, both categories have a supporter in $M$, so both rules return $P$.

  \item Choose distinct $i,j\in M$ and distinct $r,s,t\in P$.
  By \cref{lem:real}, choose a profile where $i$ assigns $x$ to $r$, $j$ assigns it to $s$, and everyone assigns it to either $r$ or $s$.
  The outcome is $P$, so it includes the unsupported category $t$.
\end{enumerate}
\par\nobreak\noindent\hfill$\blacksquare$\par

\subsection*{Proof of \texorpdfstring{\cref{thm:junta-common-member}}{Proposition}}

\begin{enumerate}[label=(\alph*),leftmargin=*,noitemsep]
  \item Suppose that \cref{eq:pair-nonempty,eq:pair-coverage} hold.
  Fix a profile $\cl\in\C^N$.
  For each $x$, choose $i\in\bigcap_{r\in P}M_{x,r}$.
  Then $i\in M_{x,c_i(x)}$, so $x$ receives $c_i(x)$.
  For each $r$, choose $i\in\bigcap_{x\in X}M_{x,r}$.
  Surjectivity gives an object $x$ with $c_i(x)=r$, and $i\in M_{x,r}$ ensures that $x$ receives $r$.
  Thus every object receives a category and every category is used.

  \item Conversely, suppose the rule is a CAC. If \cref{eq:pair-nonempty} fails at $x$, each $i$ can choose $r_i$ with $i\notin M_{x,r_i}$.
  By \cref{lem:real}, realise the assignments $c_i(x)=r_i$.
  No category has a supporter in its junta at $x$, leaving $x$ without a category.

  \item Likewise, if \cref{eq:pair-coverage} fails for $r$, each $i$ can choose $y_i$ with $i\notin M_{y_i,r}$.
  Set $c_i(y_i)=r$ and map the remaining objects onto $P\setminus\{r\}$, possible because $m-1\geq p-1$.
  Then $c_i$ is surjective and uses $r$ only at $y_i$.
  Every supporter of $r$ is therefore outside the relevant junta, leaving $r$ unused.
\end{enumerate}
The special cases follow by substitution.
\par\nobreak\noindent\hfill$\blacksquare$\par

\subsection*{Proof of \texorpdfstring{\cref{thm:pair-representation}}{Theorem}}

\begin{enumerate}[label=(\alph*),leftmargin=*,noitemsep]
  \item Suppose $F$ satisfies Independence, Supportiveness and Responsibility. For each $x,r$, let $M_{x,r}$ consist of the individuals whose assignment of $x$ to $r$ guarantees its acceptance at every profile.
  By Supportiveness and \cref{lem:local-responsibility}, every accepted assignment has a supporter in $M_{x,r}$.
  Conversely, the support of any member of $M_{x,r}$ ensures acceptance by definition.
  Thus $r\in F(\cl)(x)\Longleftrightarrow S_r(x,\cl)\cap M_{x,r}\neq\emptyset$.
  This gives \cref{eq:pair-representation}, and \cref{thm:junta-common-member} gives \cref{eq:pair-nonempty,eq:pair-coverage}.

  \item \textit{Uniqueness.} By \cref{lem:real}, any individual $i$ can be the sole supporter of $r$ at $x$.
  At such a profile, $r$ is accepted exactly when $i\in M_{x,r}$, which determines every $M_{x,r}$ uniquely.

  \item Conversely, let $F$ have the stated junta representation. The rule is a CAC by \cref{thm:junta-common-member} and satisfies Independence and Supportiveness by its definition.
  For every accepted assignment, choose $i\in S_r(x,\cl)\cap M_{x,r}$.
  Keeping $c_i$ unchanged preserves this support whatever the others report, so Responsibility holds.
\end{enumerate}
\par\nobreak\noindent\hfill$\blacksquare$\par

\subsection*{Proof of \texorpdfstring{\cref{cor:FRR-junta}}{Corollary}}

\begin{enumerate}[label=(\roman*),leftmargin=*,noitemsep]
  \item By \cref{thm:pair-representation}, write $F=J^{(M_{x,r})}$.
  Fix $x$, distinct $r,s$, and $i$.
  By \cref{lem:real}, take a profile where $i$ alone supports $r$ at $x$.
  Swapping $r$ and $s$ makes $i$ the sole supporter of $s$ at $x$.
  Category Neutrality therefore gives $i\in M_{x,r}\Longleftrightarrow i\in M_{x,s}$.
  Hence $M_{x,r}=M_x$ for every $r$, and \cref{thm:junta-common-member} gives \cref{eq:common-member}.
  Conversely, these rules satisfy Category Neutrality and the axioms of \cref{thm:pair-representation}.

  \item By (i), write $F=J^{(M_x)}$.
  Take a profile where $i$ alone supports $r$ at $x$, and swap objects $x$ and $y$ in every report.
  Now $i$ alone supports $r$ at $y$, so Object Neutrality gives $i\in M_x$ if and only if $i\in M_y$.
  Thus all $M_x$ equal a common nonempty coalition $M$.
  Conversely, a uniform junta rule also satisfies Object Neutrality.
\end{enumerate}
\par\nobreak\noindent\hfill$\blacksquare$\par

\subsection*{Proof of \texorpdfstring{\cref{thm:general_junta}}{Theorem}}

\begin{enumerate}[label=(\alph*),leftmargin=*,noitemsep]
  \item Suppose $F$ satisfies Independence, Supportiveness and Split Stability, and fix $x$.
  By Independence, write $f_x(\cl_x)=F(\cl)(x)$.
  By \cref{lem:rich}, each opinion vector can be realised with every individual's report surjective on $X\setminus\{x\}$.
  All changes at $x$ below therefore preserve surjectivity.

  \begin{claim}\label{claim:junta_for_k}
    Whether $k\in f_x(\pi)$ depends only on $S_k(\pi)$.
  \end{claim}
  \begin{proof}\renewcommand{\qedsymbol}{$\square$}
    Fix $S\subseteq N$ and $\ell\neq k$.
    Start with the opinion vector assigning $k$ to members of $S$ and $\ell$ to everyone else.
    To reach any $\pi$ with $S_k(\pi)=S$, move each nonempty group $S_r(\pi)$, $r\notin\{k,\ell\}$, from $\ell$ to $r$ in turn.
    Each destination is vacant, so \ref{cond:SI-i} preserves whether $k$ is accepted throughout.
    \par\nobreak\noindent\hfill\qedhere\par
  \end{proof}

  Let $\W_x(k)$ be the family of support sets that lead to acceptance of $k$.
  Supportiveness and \cref{lem:S-implies-U} give $\emptyset\notin\W_x(k)$ and $N\in\W_x(k)$.

  \begin{claim}\label{claim:junta-across_categories}
    The family $\W_x(k)$ is the same for every category $k$.
  \end{claim}
  \begin{proof}\renewcommand{\qedsymbol}{$\square$}
    Choose distinct $k,k',\ell$ and a nonempty $S\subseteq N$.
    Let members of $S$ support $k$ and everyone else support $\ell$.
    Move all of $S$ from $k$ to the vacant category $k'$.
    By \ref{cond:SI-ii}, $k$ was accepted exactly when $k$ or $k'$ is now accepted.
    Supportiveness excludes the now unsupported $k$, so $S\in\W_x(k)$ if and only if $S\in\W_x(k')$.
    The empty set belongs to neither family.
    \par\nobreak\noindent\hfill\qedhere\par
  \end{proof}

  Writing $\W_x$ for this common family, we have
  \begin{equation}
    \label{eq:Wx-representation}
    F(\cl)(x)=\{k\in P\mid S_k(x,\cl)\in\W_x\}.
  \end{equation}

  \begin{claim}\label{claim:grill-disjoint_sets-iff}
    For disjoint $A,B\subseteq N$, $A\cup B\in\W_x$ if and only if $A\in\W_x$ or $B\in\W_x$.
  \end{claim}
  \begin{proof}\renewcommand{\qedsymbol}{$\square$}
    If either set is empty, use $\emptyset\notin\W_x$.
    Otherwise, choose distinct $k,k',\ell$.
    Let $A\cup B$ support $k$ and everyone else support $\ell$, then move $B$ to the vacant category $k'$.
    The two support sets become $A$ and $B$, so \ref{cond:SI-ii} gives the claim.
    \par\nobreak\noindent\hfill\qedhere\par
  \end{proof}

  The claim implies upward closure: if $A\in\W_x$ and $A\subseteq T$, apply it to $T=A\cup(T\setminus A)$.
  It also gives the grill property: if $A\cup B\in\W_x$, apply it to $A\cup(B\setminus A)$ and use upward closure.
  Thus $\W_x$ is a grill.
  Since $N$ is finite, \cref{prop:grill-structure} gives a unique nonempty $M_x$ with $\W_x=\G_{M_x}$.
  Equation~\eqref{eq:Wx-representation} then gives $F=J^{(M_x)}$, and \cref{thm:junta-common-member} gives \cref{eq:common-member}.

  \item Conversely, let $F=J^{(M_x)}$ satisfy \cref{eq:common-member}.
  It is a CAC by \cref{thm:junta-common-member}; its definition gives Independence and Supportiveness.
  Consider an admissible defection of $B\subseteq S_k(x,\cl)$ to a vacant category $k'$.
  Other support sets are unchanged, giving \ref{cond:SI-i}.
  The support of $k$ splits into $S_k(x,\cl)\setminus B$ and $B$.
  At least one part meets $M_x$ exactly when the original set does, giving \ref{cond:SI-ii}.
\end{enumerate}
\par\nobreak\noindent\hfill$\blacksquare$\par

\subsection*{Proof of \texorpdfstring{\cref{cor:uniform_junta}}{Corollary}}

By \cref{thm:general_junta}, write $F=J^{(M_x)}$.
Swapping objects as in the proof of \cref{cor:FRR-junta}(ii) shows that Object Neutrality makes all $M_x$ equal.
Conversely, every uniform junta rule satisfies Object Neutrality and the axioms of \cref{thm:general_junta}.
\par\nobreak\noindent\hfill$\blacksquare$\par

\subsection*{Proof of \texorpdfstring{\cref{thm:frrc-characterisation}}{Theorem}}

On finite populations, Responsibility and Rejection Stability are equivalent under Supportiveness by \cref{lem:FRR-FRRC}.
Parts (i) and (ii) therefore follow from \cref{thm:pair-representation,cor:FRR-junta}, respectively.
\par\nobreak\noindent\hfill$\blacksquare$\par

\subsection*{Proof of \texorpdfstring{\cref{prop:una-replace}}{Proposition}}

In each case, we show that a category with no supporters is rejected.
Fix $\cl,x,r$ with $S_r(x,\cl)=\emptyset$.
\begin{enumerate}[label=(\roman*),leftmargin=*,noitemsep]
  \item Let $k_1,\ldots,k_q$ be the categories used at $x$ in $\cl$.
  By \cref{lem:rich}, start from a profile where everyone assigns $x$ to $k_1$ and every report is surjective on $X\setminus\{x\}$.
  Unanimity rejects $r$.
  For $j=2,\ldots,q$, move $S_{k_j}(x,\cl)$ from $k_1$ to the vacant category $k_j$.
  Every move is admissible, and \ref{cond:SI-i} preserves rejection of $r$ because $r\notin\{k_1,k_j\}$.
  The resulting opinion vector is $\cl_x$, so Independence gives $r\notin F(\cl)(x)$.

  \item By \cref{lem:real}, take a profile $\bm d$ where everyone assigns $x$ to some $s\neq r$.
  Unanimity rejects $r$ there.
  Apply Rejection Stability with both source profiles equal to $\bm d$ and target $\cl$.
  Its report condition is vacuous because $\cl$ has no supporter of $r$ at $x$, so $r\notin F(\cl)(x)$.
\end{enumerate}
\par\nobreak\noindent\hfill$\blacksquare$\par

\subsection*{Proof of \texorpdfstring{\cref{thm:limiting}}{Theorem}}

\begin{lemma}\label{lem:grill}
  Let $N$ be finite or infinite, and let $F$ satisfy Independence, Supportiveness, and either Split Stability (with $m>p\geq3$) or Rejection Stability (with $m\geq p\geq2$).
  Then, for every $x\in X$ and $r\in P$:
  \begin{enumerate}[label=(\roman*),leftmargin=*,noitemsep]
    \item every $A\subseteq N$ is either winning or null for $r$ at $x$;
    \item $\W_{x,r}$ is a grill.
  \end{enumerate}
  Under Split Stability, or under Rejection Stability and Category Neutrality, $\W_{x,r}=\W_x$ does not depend on $r$; under Object Neutrality as well, it does not depend on $x$.
\end{lemma}

\begin{proof}\leavevmode
  \begin{enumerate}[label=(\alph*),leftmargin=*,noitemsep]
    \item \textit{Split Stability.}
    The three claims in the proof of \cref{thm:general_junta} and the argument following them give the result.
    Those steps apply to infinite populations as well: the number of defections is finite because $P$ is finite.

    \item \textit{Rejection Stability.} Fix $x,r$.
    Suppose $\bm d$ rejects $r$ with support set $A$, and take any profile $\bm e$ with the same support set.
    Form $\cl$ by taking $e_i$ for $i\in A$ and $d_i$ for everyone else.
    Since $\cl_x=\bm d_x$, Independence makes $\cl$ reject $r$.
    Every supporter in $\bm e$ retains her report from $\cl$, so Rejection Stability, with both source profiles equal to $\cl$, makes $\bm e$ reject $r$.
    Thus every coalition that is not winning is null.

    Let $\I_{x,r}$ be the family of null coalitions.
    Take $A,B\in\I_{x,r}$ and any profile $\bm e$ with support set $S\subseteq A\cup B$.
    Choose profiles $\cl,\bm d$ with support sets $A,B$, retaining $e_i$ in $\cl$ for $i\in S\cap A$ and in $\bm d$ for $i\in S\cap B$.
    Complete the other reports using \cref{lem:real}.
    Both profiles reject $r$, and each supporter in $\bm e$ retains her report from at least one of them.
    Rejection Stability therefore makes $\bm e$ reject $r$, so $S$ is null.
    Hence $\I_{x,r}$ is downward closed and closed under union.
    Supportiveness and \cref{lem:S-implies-U} give $\emptyset\in\I_{x,r}$ and $N\notin\I_{x,r}$.
    Its complement $\W_{x,r}$ is therefore a grill.

    \item \textit{Neutrality.} Swapping $r$ and $s$ turns each support set for $r$ into the same support set for $s$, so Category Neutrality gives $\W_{x,r}=\W_{x,s}$.
    Likewise, swapping $x$ and $y$ and applying Object Neutrality gives $\W_{x,r}=\W_{y,r}$.
  \end{enumerate}
  \par\nobreak\noindent\hfill\qedhere\par
\end{proof}

\begin{lemma}\label{lem:limiting-cac}
  Let $(\mathcal M_x)_{x\in X}$ be a family of directed families, and write $\G(\mathcal M_x)=\{S\subseteq N\mid S\cap M\neq\emptyset\text{ for every }M\in\mathcal M_x\}$.
  \begin{enumerate}[label=(\roman*),noitemsep]
    \item Each $\G(\mathcal M_x)$ is a grill, and $r\in L^{(\mathcal M_x)}(\cl)(x)$ if and only if $S_r(x,\cl)\in\G(\mathcal M_x)$.
    \item Conversely, for every grill $\G$, its family of blocking coalitions $\G^\ast$ is directed and $\G=\G(\G^\ast)$.
    \item $L^{(\mathcal M_x)}$ is a CAC if and only if \cref{eq:selection} holds. This condition holds automatically for uniform limiting junta rules.
  \end{enumerate}
\end{lemma}

\begin{proof}\leavevmode
  \begin{enumerate}[label=(\roman*),leftmargin=*,noitemsep]
    \item Since $\mathcal M_x$ is a nonempty family of nonempty coalitions, $N\in\G(\mathcal M_x)$ and $\emptyset\notin\G(\mathcal M_x)$.
    Upward closure follows from the definition.
    If $A,B\notin\G(\mathcal M_x)$, choose $M_1,M_2\in\mathcal M_x$ disjoint from $A,B$, respectively.
    Directedness gives $M_3\in\mathcal M_x$ with $M_3\subseteq M_1\cap M_2$.
    Then $M_3$ is disjoint from $A\cup B$, proving the grill property.
    The representation of $L^{(\mathcal M_x)}$ is its definition.

    \item By \cref{lem:duality}, $\G^\ast$ is a filter, hence directed, and $\G=\G(\G^\ast)$.

    \item At each object, the finitely many support sets partition $N$.
    The grill property therefore makes at least one of them winning, so the outcome is nonempty.
    It remains to check that every category is used.

    Suppose \cref{eq:selection} holds and $r$ is unused.
    For each $x$, choose $M_x\in\mathcal M_x$ disjoint from $S_r(x,\cl)$.
    Any $i\in\bigcap_xM_x$ would then assign no object to $r$, contradicting surjectivity.

    Conversely, suppose $M_x\in\mathcal M_x$ for each $x$ and $\bigcap_xM_x=\emptyset$.
    Fix $r$ and, for each $i$, choose $y_i$ with $i\notin M_{y_i}$.
    Set $c_i(y_i)=r$ and map the remaining objects onto $P\setminus\{r\}$, possible because $m-1\geq p-1$.
    Thus $S_r(x,\cl)\cap M_x=\emptyset$ at every $x$, leaving $r$ unused.

    For a uniform rule, directedness and finiteness of $X$ ensure that every intersection $\bigcap_xM_x$ contains a member of $\mathcal M$.
    That member is nonempty, so \cref{eq:selection} holds.
  \end{enumerate}
  \par\nobreak\noindent\hfill\qedhere\par
\end{proof}

\begin{proof}[Proof of \cref{thm:limiting}]\leavevmode
  \begin{enumerate}[label=(\alph*),leftmargin=*,noitemsep]
    \item Suppose $F$ satisfies either set of axioms. By \cref{lem:grill}, acceptance at $x$ is determined by a grill $\W_x$ common to all categories.
    Set $\mathcal M_x=\W_x^\ast$.
    By \cref{lem:limiting-cac}(ii), $\mathcal M_x$ is directed and $\W_x=\G(\mathcal M_x)$, so $F=L^{(\mathcal M_x)}$.
    Part (iii) gives \cref{eq:selection} because $F$ is a CAC.
    Under Object Neutrality, \cref{lem:grill} makes $\W_x$, and hence $\mathcal M_x$, the same for every object.

    \item Conversely, let $F=L^{(\mathcal M_x)}$ satisfy \cref{eq:selection} and write $\G_x=\G(\mathcal M_x)$.
    By \cref{lem:limiting-cac}, $F$ is a CAC with $r\in F(\cl)(x)\Longleftrightarrow S_r(x,\cl)\in\G_x$.
    This gives Independence and Category Neutrality; $\emptyset\notin\G_x$ gives Supportiveness.

    For Split Stability, an admissible defection to a vacant category leaves all other support sets unchanged, giving \ref{cond:SI-i}.
    It splits the original support set into two parts.
    The grill property and upward closure say that the original set is winning exactly when at least one part is winning, giving \ref{cond:SI-ii}.

    For Rejection Stability, let $\cl,\bm d$ reject $r$ at $x$ and let $\bm e$ satisfy its report condition.
    Then $S_r(x,\bm e)\subseteq S_r(x,\cl)\cup S_r(x,\bm d)$.
    The union is losing by the grill property, and every subset of a losing set is losing by upward closure.
    Thus $\bm e$ rejects $r$ at $x$.
    Uniform limiting junta rules also satisfy Object Neutrality.
  \end{enumerate}
  \par\nobreak\noindent\hfill\qedhere\par
\end{proof}

\subsection*{Proof of \texorpdfstring{\cref{prop:lj-vs-junta}}{Proposition}}

\begin{enumerate}[label=(\roman*),leftmargin=*,noitemsep]
  \item If $M_x$ is the smallest member of $\mathcal M_x$, a coalition meets every member exactly when it meets $M_x$.
  Hence $L^{(\mathcal M_x)}=J^{(M_x)}$.

  Conversely, suppose $L^{(\mathcal M_x)}=J^{(M'_x)}$.
  By \cref{lem:real}, every coalition can occur as a support set at $x$, so $\G(\mathcal M_x)=\G_{M'_x}$.
  For each $i\in M'_x$, the singleton $\{i\}$ is winning and therefore meets every member of $\mathcal M_x$.
  Thus $M'_x$ is contained in every member.
  Since $N\setminus M'_x$ is losing, some $M\in\mathcal M_x$ is disjoint from it, so $M\subseteq M'_x$.
  Hence $M=M'_x$ is the smallest member.

  \item Since $N$ is infinite, cofinite coalitions are nonempty and closed under finite intersections, hence directed.
  A finite set $S$ misses the cofinite set $N\setminus S$, while an infinite set meets every cofinite set.
  Thus a category is accepted exactly when its support set is infinite.

  Suppose this rule were $J^{(M_{x,r})}$.
  Fix $x,r$.
  Unanimous support is accepted, so $M_{x,r}\neq\emptyset$.
  For $i\in M_{x,r}$, \cref{lem:real} gives a profile where $i$ alone supports $r$ at $x$.
  The junta rule accepts $r$ there, whereas the limiting rule rejects it because its support set is finite.
  This is a contradiction.
\end{enumerate}
\par\nobreak\noindent\hfill$\blacksquare$\par

\section{Independence of the axioms}\label{app:independence}

The following CACs show that no axiom can be dropped from the main characterisations.
We assume that $N$ has at least two members, and fix distinct categories $a,b\in P$ and an object $x_0\in X$ where needed.

\begin{table}[ht]
  \centering
  \caption{Examples showing the independence of the axioms. The examples $E_1$, $E_2$, and $E_5$ require $m>p$, $m\geq3$ or $p\geq3$, and $n\geq3$, respectively.}
  \label{tab:independence}
  \small
  \begin{tabular}{lcccc}
    \toprule
    Characterisation & w/o Independence & w/o Supportiveness & w/o third axiom & w/o neutrality \\
    \midrule
    \cref{thm:pair-representation} (Resp.) & $E_2$ & $E_3$ & $E_4$ & -- \\
    \cref{cor:FRR-junta}(i) (Resp., CN) & $E_1$ & $E_3$ & $E_5$ & $E_6$ \\
    \cref{thm:general_junta} (SS) & $E_1$ & $E_3$ & $E_4$ & -- \\
    \cref{cor:uniform_junta} (SS, ON) & $E_1$ & $E_3$ & $E_5$ & $E_7$ \\
    \cref{thm:frrc-characterisation}(i) (Rej. Stab.) & $E_2$ & $E_3$ & $E_4$ & -- \\
    \cref{thm:frrc-characterisation}(ii) (Rej. Stab., CN) & $E_1$ & $E_3$ & $E_5$ & $E_6$ \\
    \bottomrule
  \end{tabular}
\end{table}

\begin{itemize}
  \item[$E_1$.] (Requires $m>p$.) $E_1(\cl)(x)=\{c_1(x)\}\cup\{c_2(x)\mid c_2(y)=c_2(x)\text{ for some }y\neq x\}$.
  It satisfies Supportiveness, Responsibility, Rejection Stability, Split Stability and both neutrality axioms, but not Independence (\cref{subsec:FRRC}).
  This is the rule $F^3$ of \cref{ex:running-split}.
  For Split Stability, note that in an admissible defection individual~$2$ can defect from $k$ only if $k$ appears elsewhere in $c_2$, and the vacant category $k'$ appears elsewhere in $c_2$ because $c_2$ is surjective; hence individual~$2$ is included before the defection if and only if she is included after it, and the outcome changes only within $\{k,k'\}$.
  \item[$E_2$.] (Requires $m\geq3$ or $p\geq3$.) For fixed $y_0\in X$ and $r_0\in P$, $E_2(\cl)(x)=\{c_1(x)\}\cup\{c_2(x)\mid c_2(y_0)=r_0\}$.
  It satisfies Supportiveness, Responsibility and Rejection Stability, because each accepted category is secured by the individual classification of individual~$1$ or~$2$, but not Independence.
  \item[$E_3$.] The constant CAC $E_3(\cl)(x)=P$ satisfies Independence, Split Stability, Responsibility, Rejection Stability and both neutrality axioms, but not Supportiveness.
  \item[$E_4$.] $E_4(\cl)(x)=\{c_i(x)\mid i\in N\}$ for $x\neq x_0$, and $E_4(\cl)(x_0)=\{b\}$ if every individual assigns $x_0$ to $b$ and $E_4(\cl)(x_0)=\{c_i(x_0)\mid i\in N\}\setminus\{b\}$ otherwise.
  It is a CAC: category $b$ is used at some object other than $x_0$ unless every individual assigns $b$ only to $x_0$, in which case the profile is unanimous on $b$ at $x_0$.
  It satisfies Independence and Supportiveness.
  It violates Responsibility at a profile that is unanimous on $b$ at $x_0$, hence also Rejection Stability by \cref{lem:FRR-FRRC}, and it violates Split Stability when $p\geq3$: if some individuals assign $x_0$ to $a$ and the others to $b$, letting all supporters of $b$ defect to a vacant category $k'$ changes the outcome from $\{a\}$ to $\{a,k'\}$, contrary to \ref{cond:SI-ii}.
  \item[$E_5$.] (Requires $n\geq3$.) $E_5(\cl)(x)=\{c_1(x)\}\cup\{r\in P\mid|S_r(x,\cl)|\geq2\}$.
  It satisfies Independence, Supportiveness and both neutrality axioms, but not Responsibility, and hence not Rejection Stability.
  \item[$E_6$.] $J^{(M_{x,r})}$ with $M_{x,a}=\{1,2\}$ and $M_{x,r}=\{1\}$ for $r\neq a$ satisfies Independence, Supportiveness, Responsibility and Rejection Stability, but not Category Neutrality.
  \item[$E_7$.] $J^{(M_x)}$ with $M_{x_0}=\{1,2\}$ and $M_x=\{1\}$ for $x\neq x_0$ satisfies all axioms of \cref{thm:general_junta,cor:FRR-junta} except Object Neutrality.
\end{itemize}

\end{document}